\documentclass[12pt]{article}

\pdfoutput=1
\usepackage{color}
\usepackage{epsfig, palatino}
\usepackage{pstricks,pst-node,pst-tree}
\usepackage{epic}
\usepackage{mathrsfs}
\usepackage{ae} 
\usepackage[T1]{fontenc}
\usepackage[ansinew]{inputenc}
\usepackage{amsmath}
\usepackage{amssymb}
\usepackage{graphicx}
\usepackage{ulem}
\definecolor{darkblue}{cmyk}{0.9,0.9,0,0}
\definecolor{darkgreen}{cmyk}{0.9,0,0.9,0}

\definecolor{blueblue}{cmyk}{0.73,0.28,0,0.5}
\definecolor{lightblue}{RGB}{55,171,200}
\definecolor{grey}{gray}{0.55}

\definecolor{pink}{cmyk}{0., 0.9859943977591037, 0.3571428571428571, 0.16000000000000003}
\definecolor{lightpink}{cmyk}{0., 0.5, 0.5, 0.}
\definecolor{lightgreen}{cmyk}{0.24175824175824182, 0., 0.9615384615384616, 0.28627450980392155}

\usepackage[colorlinks=true,linkcolor=darkblue,citecolor=darkblue,urlcolor=darkblue]{hyperref}
\usepackage{cite}
\usepackage{hyperref}
\usepackage{wasysym}
\usepackage{varioref}
\usepackage{makeidx}
\usepackage[english]{babel}
\usepackage{simplewick}
\usepackage{array}
\usepackage{multirow}
\usepackage{tabularx}
\usepackage[font={small}]{caption}

\usepackage{ulem}

\newcommand{\calc}{\mbox{${\cal C}$}}

\renewcommand{\Re}{\ensuremath{\mathrm{Re}}}
\renewcommand{\Im}{\ensuremath{\mathrm{Im}}}
\newcommand{\half}{\ensuremath{\frac{1}{2}}}

\def\({\left(}
\def\){\right)}
\def\[{\left[}
\def\]{\right]}
\def\<{\langle}
\def\>{\rangle}

\newcommand{\la}[1]{\label{#1}} 

\newcommand{\beq}{\begin{equation}}
\newcommand{\eeq}{\end{equation}}
\newcommand{\beqq}{\begin{equation*}}
\newcommand{\eeqq}{\end{equation*}}
\newcommand\beqa{\begin{eqnarray}}
\newcommand\eeqa{\end{eqnarray}}

\newcommand{\nn}{\nonumber}

\makeindex

\begin{document}

\thispagestyle{empty}

\renewcommand{\thefootnote}{\fnsymbol{footnote}}
\setcounter{page}{1}
\setcounter{footnote}{0}
\setcounter{figure}{0}
\begin{center}
$$$$
{\Large\textbf{\mathversion{bold}
The O(N) S-matrix Monolith}\par}

\vspace{1.0cm}

\textrm{Luc\'ia C\'ordova$^\text{\tiny 1,2}$, Yifei He$^\text{\tiny 3,4}$, Martin Kruczenski$^\text{\tiny 4,5}$ and Pedro Vieira$^\text{\tiny 1,6}$}
\\ \vspace{1.2cm}
\footnotesize{\textit{
$^\text{\tiny 1}$ Perimeter Institute for Theoretical Physics, Waterloo, Ontario N2L 2Y5, Canada\\
$^\text{\tiny 2}$ Department of Physics and Astronomy \& Guelph-Waterloo Physics Institute, University of Waterloo, Waterloo, Ontario N2L 3G1, Canada\\
$^\text{\tiny 3}$ Institut de Physique Th\'eorique, CEA Saclay, 91191 Gif-sur-Yvette, France\\
$^\text{\tiny 4}$ Department of Physics and Astronomy, Purdue University, W. Lafayette, IN 47907, USA\\
$^\text{\tiny 5}$ Purdue Quantum Science and Engineering Institute (PQSEI), Purdue University, W. Lafayette, IN 47907, USA\\
$^\text{\tiny 6}$ Instituto de F\'isica Te\'orica, UNESP, ICTP South American Institute for Fundamental Research, 01140-070, S\~ao Paulo, Brazil
}  
\vspace{4mm}
}

\par\vspace{1.5cm}

\textbf{Abstract}\vspace{2mm}

\end{center}

 We consider the scattering matrices of massive quantum field theories with no bound states and a global $O(N)$ symmetry in two spacetime dimensions. In particular we explore the space of two-to-two S-matrices of particles of mass $m$ transforming in the vector representation as restricted by the general conditions of unitarity, crossing, analyticity and $O(N)$ symmetry. We found a rich structure in that space by using convex maximization and in particular its convex dual minimization problem. At the boundary of the allowed space special geometric points such as vertices  were found to correspond to integrable models. The dual convex minimization problem provides a novel and useful approach to the problem allowing, for example, to prove that generically the S-matrices so obtained saturate unitarity and, in some cases, that they are at vertices of the allowed space.

\noindent

\setcounter{page}{1}
\renewcommand{\thefootnote}{\arabic{footnote}}
\setcounter{footnote}{0}

 \def\nref#1{{(\ref{#1})}}

\newpage

\tableofcontents

\parskip 5pt plus 1pt   \jot = 1.5ex

\newpage
\section{Introduction} 

\begin{figure}[t]
\centering
\includegraphics[width=\textwidth]{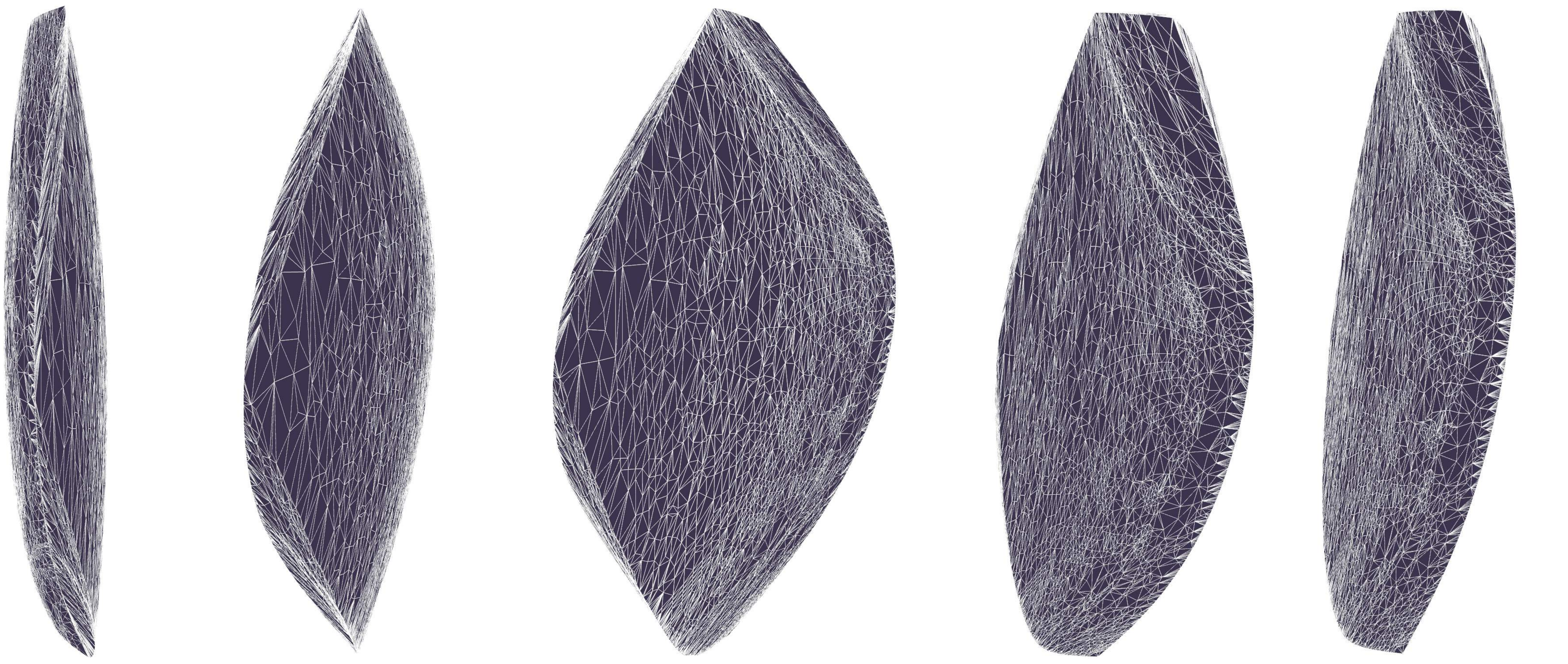}\vspace{0.3cm}
\caption{S-matrix monolith for $O(7)$ and $s_*=3m^2$. The best way to feel -- quite literally -- the various vertices, pre-vertices, edges and faces of the monolith is to 3D print it. We can easily detect nearly imperceptible vertices with one's fingertips \cite{fingertips}, see also figure \ref{lessSharp} below. We attach an ancillary file \texttt{3dPrint.stl} made out of a discretization of the monolith with more than 200,000 points (using the method of normals explained below) which can be directly printed or very efficiently visualized \cite{visualization}. To generate such 3D printing file for the convex monolith is quite simple. We generate a huge list of points belonging to the monolith and then create the convex hull of all these points using \texttt{Mathematica}'s built-in function \texttt{ConvexHullMesh} which can then be exported directly into an \texttt{.stl} file. 
}
\la{fig_monos}
\end{figure}

Consider the scattering of $O(N)$ vector particles in two dimensions in the absence of any other stable particles/bound-states as recently revisited in \cite{He:2018uxa,Cordova:2018uop}. At first sight, this looks like a harmless mathematical problem. We simply want to study the space of the three functions which have no singularities in the physical strip\footnote{We are looking at the two-to-two scattering matrix element parametrized by the center of mass energy $s$. Throughout this paper we use interchangeably $s$ or the rapidity $\theta$ defined by $s = 4m^2\cosh^2\theta/2$. In the rapidity complex plane, the physical sheet gets mapped to the strip  $0<\text{Im}(\theta)<\pi$ (see e.g. Figure 1 of \cite{Cordova:2018uop}).} $0<\text{Im}(\theta)<\pi$, are purely real when $\theta$ is purely imaginary, obey crossing and are bounded by unitarity\footnote{The unitarity conditions usually written as a constraint on the absolute value of the S-matrix elements $|S_a(\theta)|^2\leq1$ can be recast as the positive semidefinite condition below as used e.g. in \cite{Homrich:2019cbt}.}:
\beq
S_a(i\pi-\theta)=C_{ab}\,S_b(\theta)\,,\;\;\;\;\;\;\;
\left(\begin{array}{cc}
1  & S_{a}(\theta) \\
 {S}^*_{a}(\theta)  & 1 
\end{array}\right) \succeq 0\;\;\; \text{for }\theta\in\mathbb R\,,
\label{ourProblem}
\eeq
where $a$ labels the three possible representations: singlet, antisymmetric and symmetric traceless and $C_{ab}$ is the crossing matrix where the group parameter $N$ enters (for the explicit form see \eqref{Cmatrix} in appendix~\ref{notation}). That is it, this is our problem. 


The $O(N)$ S-matrix space defined through (\ref{ourProblem}) is an infinite dimensional convex space since it is an intersection of two convex spaces: an infinite dimensional hyperplane defined by crossing and the space of positive semi-definite matrices as imposed by unitarity. Throughout this paper, we use a three-dimensional section corresponding to the real values of $S_a(\theta_*)$ for various $\theta_*$ along the imaginary axis with $\text{Im}\[\theta_*\] \in [0,\pi]$ (or $s_* \in [0,4m^2]$) to visualize this infinite dimensional space. These three coordinates can be thought of as effective four point-couplings measuring the interaction strength in the theory in each of the three scattering channels. The three dimensional allowed shape hence obtained is what we call the $O(N)$ {monolith} and which we illustrate in figure~\ref{fig_monos}. If $\theta_*=i \pi/2$ we are at a crossing symmetric point and this three-dimensional shape flattens out into a two-dimensional shape which we dub the $O(N)$ \textit{slate} (see shaded region in figure~\ref{fig_2dcurveB} below) and which we study in great detail in section~\ref{sec2_slate}. 

This space turns out to be extremely rich and the S-matrices living in its boundary exhibit a large number of striking features such as Yang-Baxter factorization at some special points, some rather universal emergent periodicity (in the logarithm of the physical energy) and infinitely many resonances (showing up as poles in higher sheets), sometimes arranged in nice regular patterns, some other times organized in intricate fractal structures. We also find {vertices}, edges and faces in the boundary of this space and even some new kind of hybrid structures we dub {\textit{pre-vertices}}. Finally we find that unitarity is not only satisfied but actually saturated for any real $\theta$ at all points in this boundary except at one single point which we call \textit{the yellow point} and whose S-matrix is a constant. Throughout the following we focus on the monolith for $N>2$. The special $N=2$ case is discussed in appendix~\ref{app_N2}.

\begin{figure}[t]
\centering
\includegraphics[width=.8\textwidth]{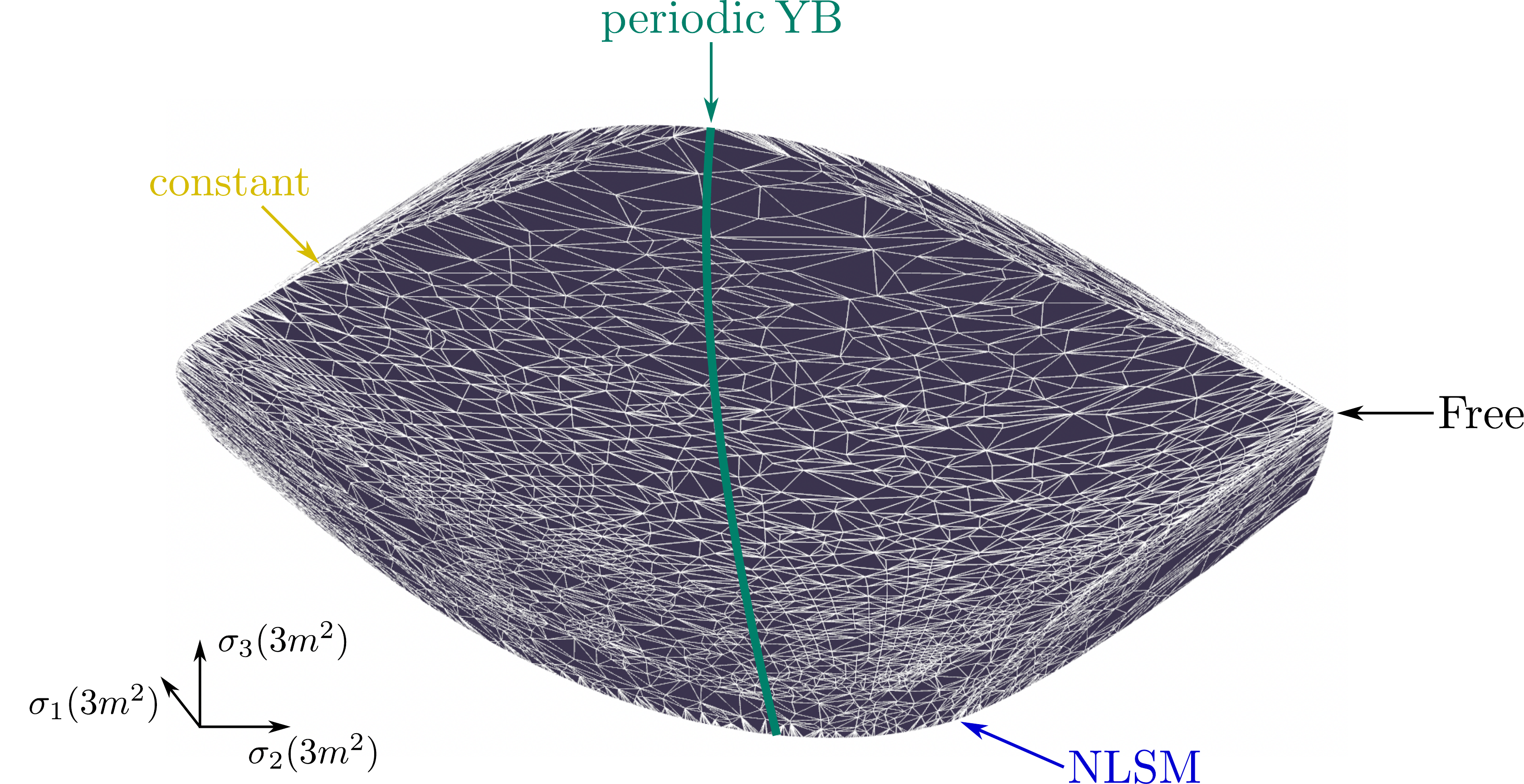}
\caption{Some features of the $O(N)$ monolith. Three arrows point to the integrable solutions corresponding to vertices (Free, NLSM) or pre-vertices (periodic YB) of the monolith. A fourth arrow points the yellow point corresponding to a constant solution which does not saturate unitarity. A line of simple (yet non-integrable) S-matrices connecting the two periodic Yang-Baxter solutions is highlighted in green. For each such special feature there is a mirror one simply related by $S_a\to -S_a$ which is a clear symmetry of the monolith.}
\la{fig_mono}
\end{figure}

Figure~\ref{fig_mono} shows some of these remarkable features. First, we highlight three integrable solutions\footnote{These three integrable S-matrices are the so called minimal solutions; multiplying these by CDD factors we obtain more integrable solutions which live inside the allowed space (and not at its boundary).}: free theory, the $O(N)$ non-linear sigma model (NLSM) and a periodic solution to the Yang-Baxter equation found in \cite{Hortacsu:1979pu} and rediscovered in \cite{Cordova:2018uop}. The first two are clear vertices of the monolith where different edges meet. For the latter the situation is more subtle since there are two edges clearly pointing towards it, but they loose their sharpness as they get closer to the integrable point, this is what we referred to as \textit{pre-vertex} before. Secondly, the \textit{yellow point} discussed above sits on one of the faces of the monolith. Notice that the space is symmetric under reflections around the origin, i.e. if we flip the sign of the S-matrix we get another viable S-matrix, so that each of the above points appears twice. Finally, there is a line on the boundary of the monolith connecting the two periodic Yang-Baxter solutions where two of the scattering channels are the same (up to a relative sign) so that the S-matrices are simple enough to write analytically (this line is explored in appendix~\ref{app_sigma2line}).



\begin{figure}[t]
\centering
\includegraphics[width=.9\textwidth]{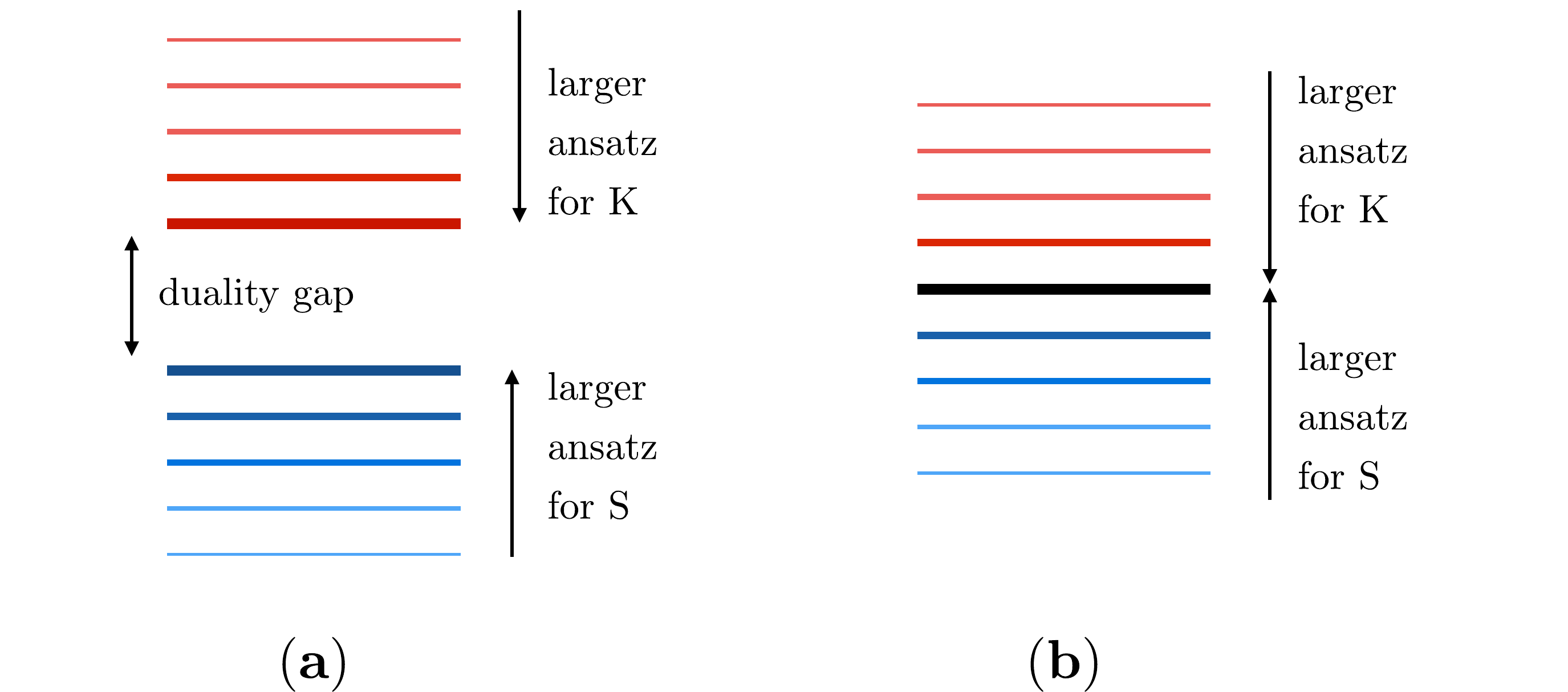}
\caption{By construction, the minimum of the dual problem (\ref{Fmin}) puts a strict upper bound on the maximum of the original problem (\ref{example}). A priori the optimal minimum and the optimal maximum are separated by what is known as the duality gap as depicted in ({\bf a}). For convex problems the duality gap is zero and thus both problems describe the very same boundary of the S-matrix space, one converging from its interior, the other from its exterior, see ({\bf b}). With different ansatze we can thus rigorously bracket the optimal bound (in black in ({\bf b})). Strictly speaking the previous statements should be qualified by the statement that both the dual and the primal problem ought to be \textit{feasible} which is the case for us.}
\la{twoDirections}
\end{figure}

How can we find the boundary of this $O(N)$ \textit{monolith} or the two-dimensional \textit{slate}? There are two natural options. The first one is to construct explicitly elements inside the space \eqref{ourProblem}.
By probing more elements in this space we obtain larger allowed regions until eventually we converge to the full $S$-matrix space. 
This is what is called the \textit{primal} problem and which has been explored in several recent S-matrix bootstrap works \cite{He:2018uxa,Cordova:2018uop,Homrich:2019cbt,Paulos:2016fap,Paulos:2017fhb,Guerrieri:2018uew,Paulos:2018fym}. 
The other option is by excluding S-matrices, that is by finding points which are outside of the S-matrix space. By excluding more and more points we should describe better and better the exterior of the S-matrix space until eventually we should converge towards the true boundary between allowed and disallowed S-matrices. This is what we call here the \textit{dual} problem. In convex optimization problems the original and dual problems usually go hand in hand; here we explore this duality in the S-matrix bootstrap context. A beautiful fact about convex optimization is that the dual and original problems should indeed converge towards the same optimal solution as depicted in the cartoon of figure \ref{twoDirections}. In our context, figure \ref{fig_2dcurveB} depicts the allowed slate space as probed through the original and dual problem. Both beautifully converge towards the very same optimal boundary (the black curve bracketed between the two blue curves).

\begin{figure}[t]
\centering
\includegraphics[width=\textwidth]{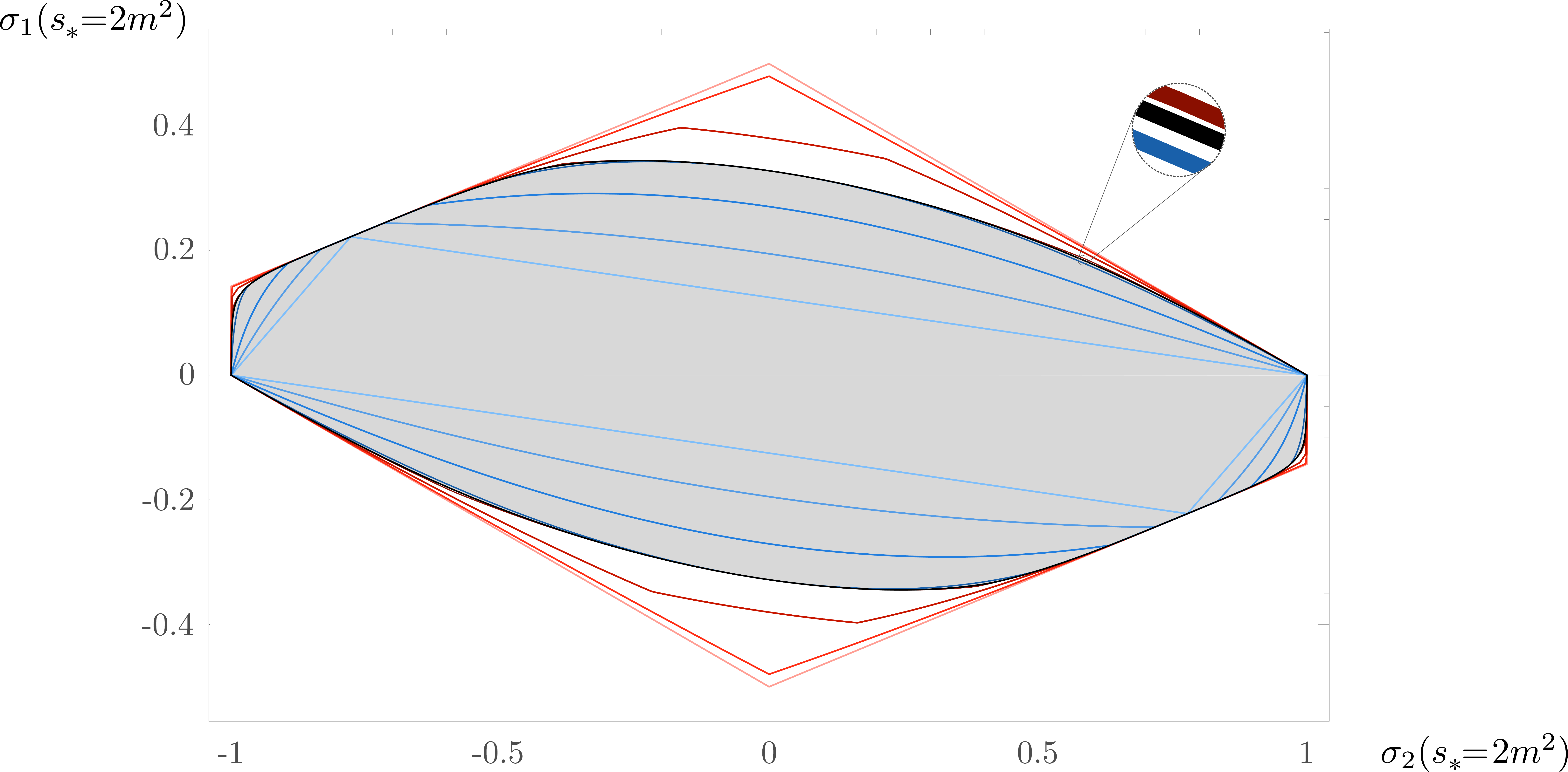}
\caption{Two-dimensional section of the monolith we call the $O(N)$ slate obtained at $s_*=2m^2$ ($\theta_*=i\pi/2$) in the $\sigma_i$ decomposition of \eqref{sigma_decomposition} for $N=7$. In black we show the optimal bound to which the primal and dual problems converge respectively from below or above. Consistent S-matrices lie on the shaded region in grey. In blue (red) we present various bounds as we take periodic ansatze in the primal (dual) problem. From lighter to darker colors we have period $\tau=0,4,6,10.25$.
}
\la{fig_2dcurveB}
\end{figure}

Let us conclude this introduction by giving some further technical details on how these problems are tackled in practice. In the primal problem in which we study directly the S-matrix space, we propose more and more general ansatze -- with several free parameters -- for smooth crossing symmetric ensembles of three functions $S_a(\theta)$.\footnote{This could be a discretized dispersion relation (free parameters would be the values of the discontinuity at a set of discrete points), a Taylor expansion (free parameters would be the Taylor coefficients), Fourier decomposition (free parameters would be the Fourier coefficients), etc.} Then we \textit{maximize} various linear functionals acting on these functions over those free parameters.

As a first example of the type of functionals used here, we can fix two components $x=S_\text{sym}(\theta_*)$ and $y=S_\text{anti}(\theta_*)$ and maximize and minimize the third component $z=S_\text{singlet}(\theta_*)$; repeating this strategy for several $(x,y)$ would yield various points on the boundary of the 3D monolith. This procedure is represented in a two-dimensional section in figure \ref{3functionals}({\bf a}). Two other functionals are more efficient. One is what we call the radial functional where we set $S_a(\theta_*)=r\, n_a$ with $n_a$ the components of a three-dimensional unit vector and we maximize $r$ to find the boundary of the monolith/slate in a particular direction $n$. This method is represented in figure \ref{3functionals}({\bf b}). Lastly, we have the so-called normal functionals where we maximize a combination $\sum_a n_a S_a(\theta_*)$. Here we find the boundary points of the S-matrix space with normal $n$, see figure \ref{3functionals}({\bf c}). This last type of functional has the advantage of putting many points close to the most interesting higher curvature regions such as vertices or edges of the S-matrix space as illustrated in figure \ref{3functionals}({\bf c}); the radial functional has the positive feature of equally populating all direction while the first type of functional has no particular advantage and, indeed, we will use it very rarely. Of course, by considering a large number of base points $(x,y)$ and many directions $n_a$ all such functionals end up describing the very same boundary. In this introduction we stick to the normal class of functionals where we maximize
\begin{figure}
\centering
\includegraphics[width=\textwidth]{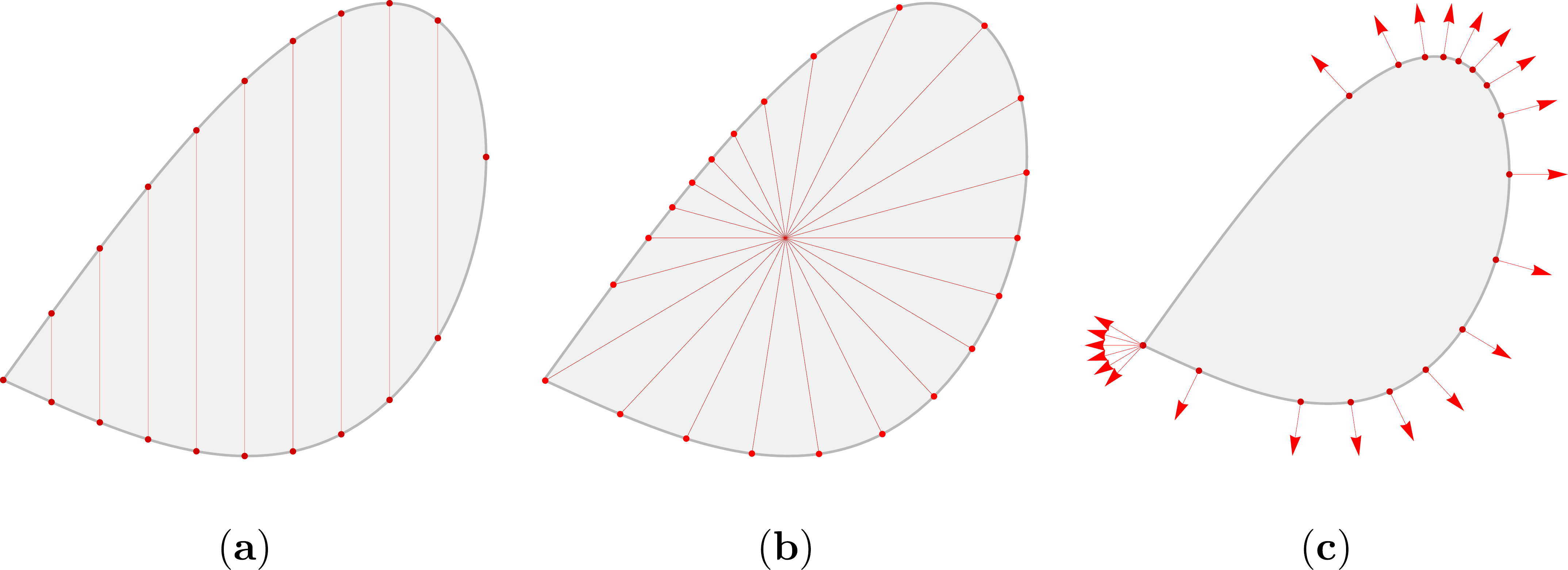}
\caption{Different maximization functionals to obtain the boundary of a certain region of the plane. In panel ({\bf a}) we fix one of the coordinates and maximize/minimize the other. In the second panel ({\bf b}) we fix a particular direction and perform a radial maximization, which useful for defining the faces of the convex space. Finally in ({\bf c}) we have the normal maximization where we have a uniform distribution of unit vectors and the maximization chooses the points where the normals are aligned with the unit vectors, resulting in a higher concentration of points in high curvature regions. }
\la{3functionals}
\end{figure}
\beq
\text{max}\Big(\mathcal{F}[S] \equiv \sum_a n_a S_a(\theta_*)\Big) \,, \qquad \text{Im}\[\theta_*\] \in[0,\pi] \la{example}
\eeq 
over crossing symmetric functionals and imposing the unitarity constraints. 
By increasing the number of free parameters describing these functions and by picking different directions $n$ we converge towards the true boundary of the S-matrix space from the inside. 

In the dual approach we reach the boundary of the S-matrix space from the outside. We start by re-writing 
\beq
\mathcal{F}
= \oint \frac{d\theta}{2\pi i}\sum_a K_a(\theta) S_a(\theta)\,,
\eeq 
which is true if each $K_a$ is a function with a pole at $\theta_*$ and residue given by $n_a$.\footnote{Strictly speaking there should also be poles at the crossing symmetric image $i\pi-\theta_*$ as explained in detail below.} The contour of integration can be taken to be a big rectangle inside the physical rapidity strip (that is the boundary of the Mandelstam physical sheet). If we impose appropriate crossing transformations on $K$ we can relate the integration over the top part of the rectangle to the bottom part so that we end up with the very same integral (times $2$) integrated over the real line alone. Since the S-matrix is at most of absolute value $1$ on the real line we conclude that 
\beq
\mathcal{F}
 \le \text{min}\Big(
\mathcal{F}_d\equiv
\sum_a \int\limits_{-\infty}^{+\infty} \frac{d\theta}{\pi} |K_a(\theta)| \la{Fmin} \Big)\,.
\eeq
We just found in this way an upper bound on the optimal solution to the primal \textit{maximization} problem (\ref{example}). We can now take an ansatz for these so far generic functions $K_a$ and solve this dual  \textit{minimization} problem. By taking 
more general ansatze for $K_a$ we get 
better estimates for the minimum of (\ref{Fmin}) which provides a sharp upper bound to the primal problem. 

As stated above, because the original problem is convex, it can be shown that this upper bound actually \textit{coincides} with the solution to the original maximization problem, see figures \ref{twoDirections} and \ref{fig_2dcurveB}. In particular, as explained in detail in section~\ref{analyticdual}, it is easy to see that this can only be true if either unitarity is saturated or the original functional is very special. This clarifies a long standing puzzle.  It was thus far stated as a mystery why was unitarity saturated at the boundary of the physical S-matrix space in many different contexts \cite{Paulos:2016but,Paulos:2017fhb,Cordova:2018uop,Homrich:2019cbt,Guerrieri:2018uew,EliasMiro:2019kyf,susy}. This dual problem, with its associated zero duality gap theorems, provides a clean explanation in the two dimensional examples.

In the rest of the paper we expand on the results mentioned in this introduction. In section~\ref{sec2_slate} we take a closer look to the space of $O(N)$ S-matrices --in particular to the two-dimensional slate-- and in section~\ref{analyticdual} we present the derivation of the dual problem and explain the bracketing procedure of figure~\ref{fig_2dcurveB}.

%
%



\section{The Monolith and Slate}\la{sec2_slate}

To approximate the infinite dimensional S-matrix space we need some clever coordinates. One possibility is to parametrize the S-matrix components by dispersion relations; two such dispersions relations were used efficiently in \cite{Cordova:2018uop} and \cite{He:2018uxa}; the code in \cite{He:2018uxa} is very fast and was the one we used to generate the heaviest plots here while the method used in \cite{Cordova:2018uop} is more reliable to explore the boundary S-matrices at large rapidities when the numerics are most challenging and was thus the one used to extract the analytic properties of the whence obtained S-matrices at various special points. Finally, a third method discussed below is to use a Fourier decomposition of the S-matrix elements; this would turn out particularly relevant due to an emergent and mysterious periodicity which the boundary S-matrices exhibit. 

\begin{figure}[t]
\centering
\includegraphics[width=\textwidth]{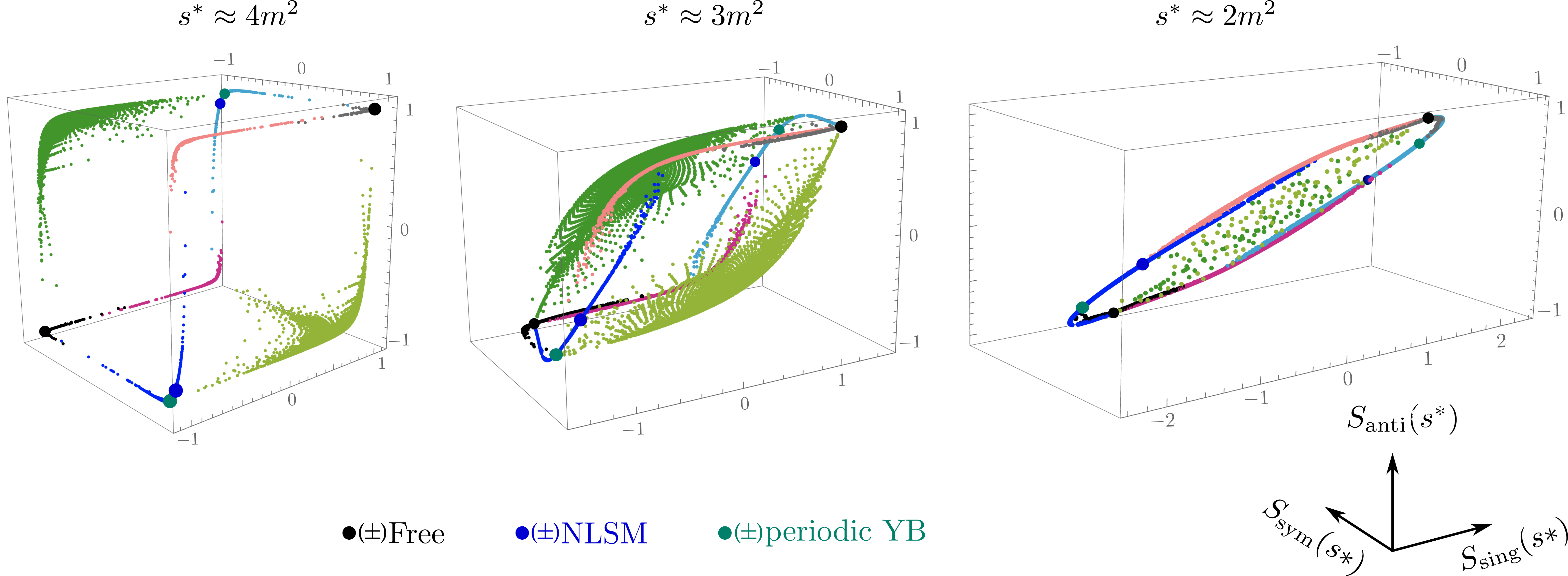}
\caption{Space of allowed S-matrices $S_a(s^*)$ for $N=7$ and different values of $s^*$, obtained using functionals of the normal type. 
Close to threshold $s^*\approx4m^2$ the space approaches the cube defined by unitarity $|S_a(s)|\leq1$ for physical values $s>4m^2$. As we decrease $s^*$, the monolith morphs into the two-dimensional slate at $s^*=2m^2$. 
At the boundary of the allowed space we have the 3 (+ 3 flipping all signs) integrable solutions, namely free theory, the NLSM \cite{Zamolodchikov:1978xm} and the periodic Yang-Baxter solution \cite{Hortacsu:1979pu}. The coloring corresponds to the associated $S_a(s=4m^2)$ values as in table~\ref{tab:colors} of appendix~\ref{app_analytic}. }
\la{fig_morph}
\end{figure}

In practice we use from a few dozens to a few hundreds of coefficients to parametrize the S-matrices. To visualize the S-matrix space, however, we need to pick a lower dimensional section as discussed in the introduction. A natural set of three variables to explore is the allowed (real) values of $S_a(s^*)$ for each of the three components for a given $s^*\in [0,4m^2]$. At the crossing symmetric point $s^*=2m^2$ these three values are no longer independent; only two are. In other words, the three-dimensional monolith flattens into a two-dimensional slate as we slide $s^*$ towards $2m^2$, see figure \ref{fig_morph}. This two dimensional slate is the simplest lower-dimensional shadow of our $O(N)$ S-matrix space. Nicely, most of the interesting kinks of the $O(N)$ space -- or at least those in the three dimensional monolith -- are still visible in this lower dimensional section which will be the main focus of this section. 


\begin{figure}[t]
\centering
\includegraphics[width=.95\textwidth]{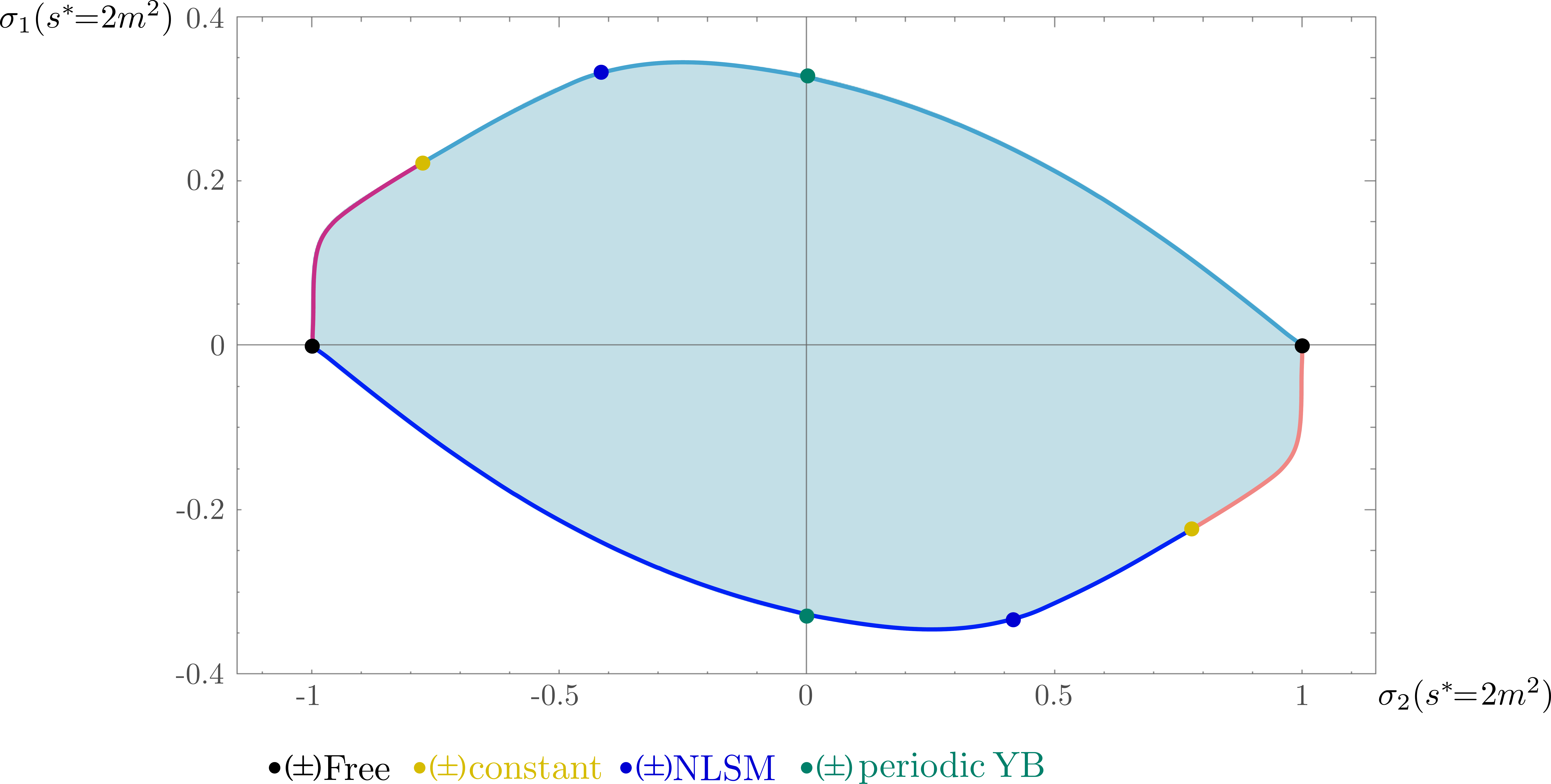}
\caption{Allowed space of S-matrices in the plane $\sigma_1(s^*=2)$ vs  $\sigma_2(s^*=2)$. The coloring at its boundary matches the convention in table~\ref{tab:colors}. We have also marked the points corresponding to known integrable S-matrices and the constant solution in \eqref{constant_soln}.}
\la{fig_2dcurve}
\end{figure}

To explore the slate we can use the primal or dual problem. Here we focus on the primal one where we give an ansatz for the S-matrices and  
maximize various functionals as discussed in the previous section. The result we obtain is represented in figure \ref{fig_2dcurve}. For each point at the boundary of this space we can extract numerically the corresponding S-matrix. Here are some remarkable features we learn from these numerics: 

\begin{figure}[t]
\centering
\includegraphics[width=\textwidth]{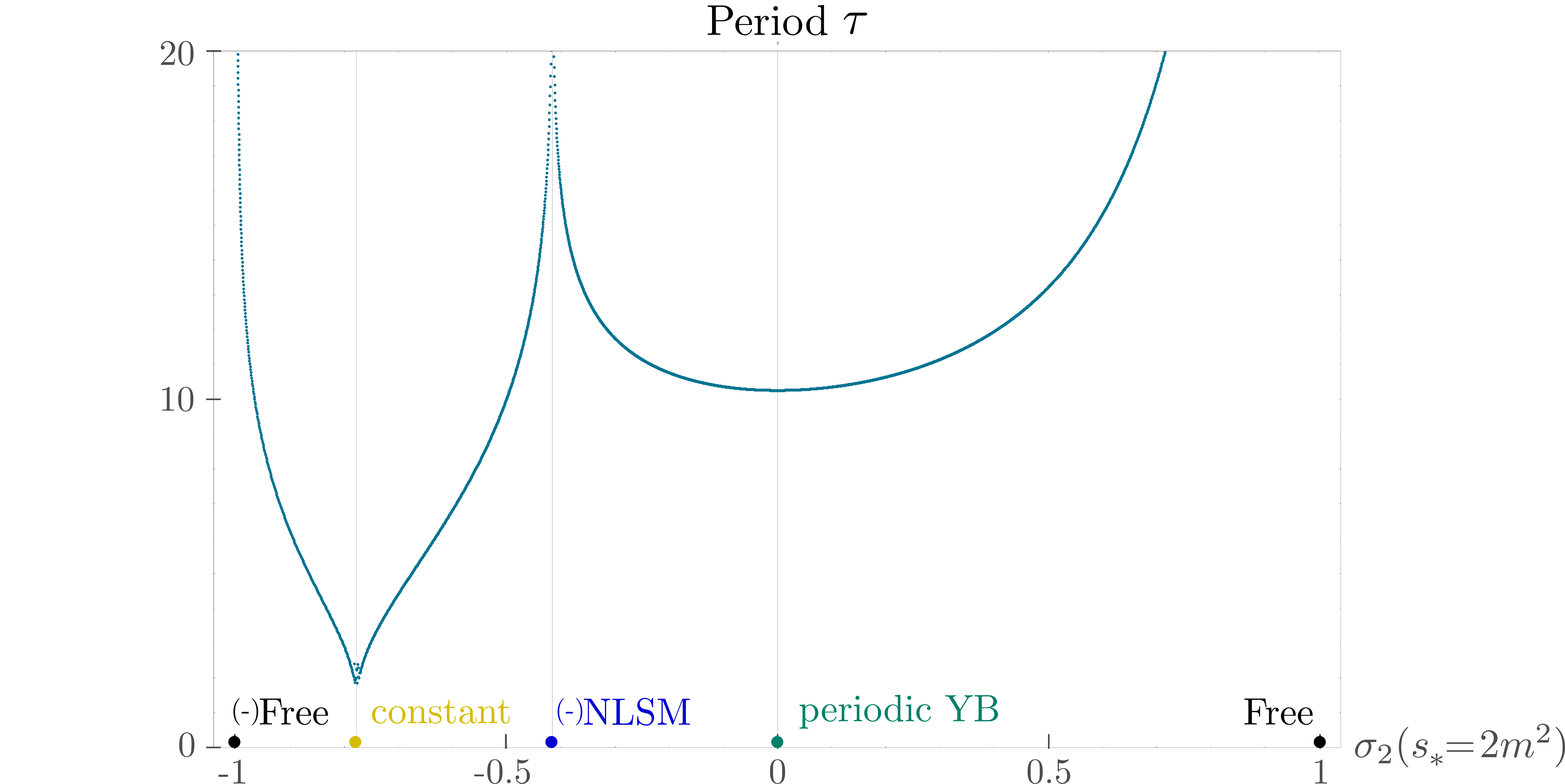}
\caption{Period $\tau$ (in the real $\theta$ direction) of the S-matrices obtained numerically along the curve defining the boundary of the  $s^*=2m^2$ plane for $N=7$. The period diverges for free theory and the non-linear sigma model, has a local minimum at the periodic Yang-Baxter solution and approaches zero for the constant solution (the plot presents some noise around the latter since the numerics have a hard time converging for small periods).
}
\la{fig_period}
\end{figure}

\begin{figure}[t]
\centering
\includegraphics[width=\textwidth]{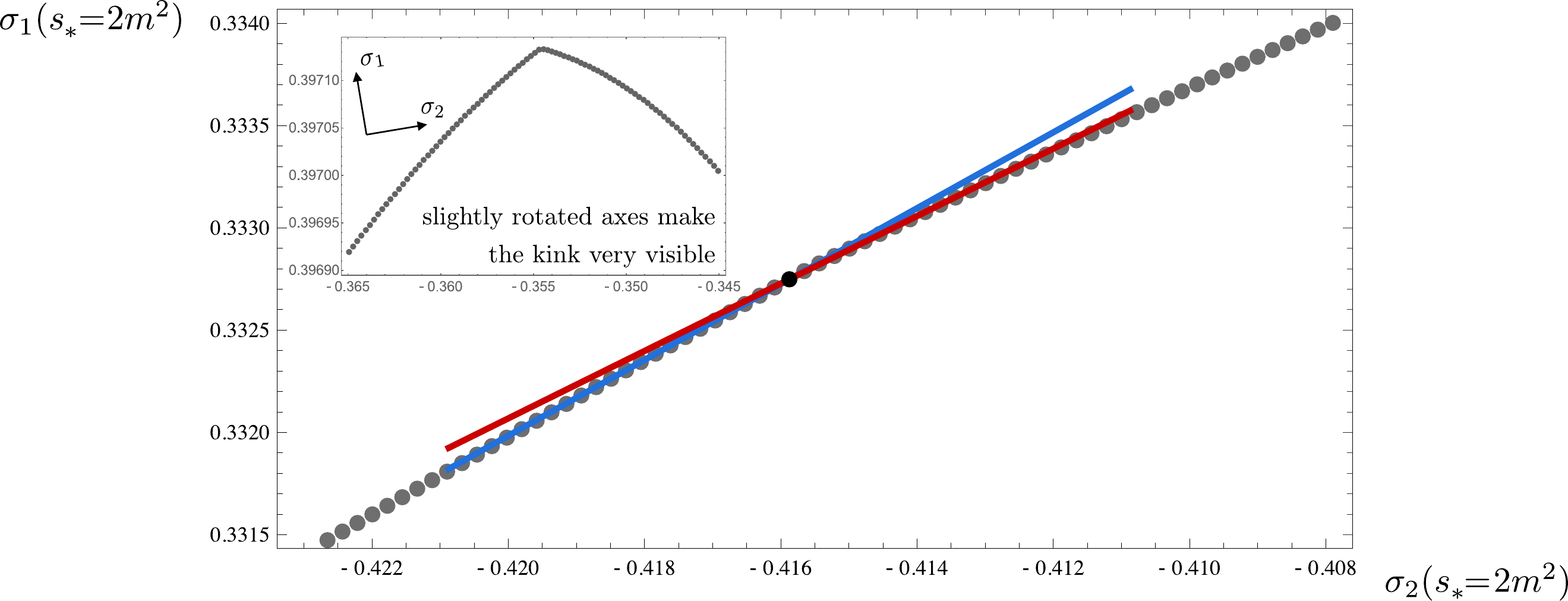}
\caption{The zoomed-in view of the less sharp kink of NLSM. The circles here are the points on the boundary of the slate and the red and blue lines are the tangents on two side of the NLSM kink at $(\sigma_2,\sigma_1)\approx (-0.4159,0.3327)$. The plot in the inset is the same as in the big one except for a simple small rotation of the axes (by $9.5$ degrees which is the approximate slope near the NLSM) which renders the NLSM kink much easier to spot.} 
\la{lessSharp}
\end{figure}

\begin{itemize}
\item A few points are special along the slate boundary: we have the free theory vertex, a less sharp kink corresponding to the $O(N)$ non-linear sigma model (NLSM) -- see figure \ref{lessSharp} --  and a point corresponding to a periodic --in real $\theta$-- integrable solution (pYB) found in \cite{Hortacsu:1979pu} and rediscovered in \cite{Cordova:2018uop}.  As mentioned in the introduction, the slate is symmetric under reflections around the origin so we get the reflected points by flipping the signs of the S-matrices.
The analytic S-matrices at these three points read $\mathbf S^\text{Free}=(1,1,1)$ and 
\beqa
\mathbf S^\text{NLSM}&=&
- \(1,\;\frac{\theta-i\pi}{\theta+i\pi},\; \frac{\theta-i\pi}{\theta+i\pi}\,\frac{\theta-i\lambda_\text{GN}}{\theta+i\lambda_\text{GN}} \)
F_{\pi+\lambda_\text{GN}}(\theta)F_{2\pi}(\theta)\la{S_NLSM} \,,\\
\la{S_NLSM}
{\bf S}^\text{pYB}&=&\,
\(\frac{\sinh\[\nu\(1-\frac{i\theta}{\pi}\)\]}{\sinh\[\nu\(1+\frac{i\theta}{\pi}\)\]},\;-1,\;1 \)
\prod\limits_{n=-\infty}^\infty F_{\pi+\frac{in\pi^2}{\nu}}(-\theta)\,,
\la{S_pYB}
\eeqa
%
where $F_a(\theta)\equiv {\Gamma \left(\frac{a+i \theta }{2 \pi }\right) \Gamma \left(\frac{a-i \theta +\pi }{2 \pi }\right)}/{\Gamma \left(\frac{a-i \theta }{2 \pi }\right) \Gamma \left(\frac{a+i \theta +\pi }{2 \pi }\right)}$, $\lambda_\text{GN}=\tfrac{2\pi}{N-2}$, $\nu=\text{arccosh}(\tfrac{N}{2})$ and we have used the notation $\mathbf S=(S_\text{sing},S_\text{anti},S_\text{sym})$.
At these three-points the S-matrix obey nice cubic factorization equations known as the Yang-Baxter equations. It is worth emphasizing that these were by no means imposed and rather come out as a mysterious outcome. It is amusing to think that had Yang-Baxter not been discovered before and these nice integrable solutions not unveiled decades ago, we could have discovered them here in these numerical explorations. 

\item Another interesting point is the yellow point between free theory and NLSM in figure~\ref{fig_2dcurve}. The S-matrix there is a simple constant solution to crossing and unitarity
\beq\la{constant_soln}
\mathbf S^\text{const}=\pm\(1,\,-1,\,\frac{N-2}{N+2}\)\,, 
\eeq
but does not obey Yang-Baxter equations. Notice that in the symmetric channel unitarity is not saturated. To our knowledge this is the first analytic solution to the S-matrix bootstrap problem where unitarity is \textit{not} saturated. We call it \textit{the yellow point}. 

If we look for constant solutions to the bootstrap problem it is actually easy to derive~(\ref{constant_soln}) analytically. First, because of crossing, all possible constant solutions lie on the same plane as the slate (i.e. must be eigenvectors of the crossing matrix). The unitarity inequalities then define a polygon on this plane which is nothing but the  innermost curve in figure~\ref{fig_2dcurveB}. 
Such polygon is simply given by $S_a=C_{ab}S_b$, $|S_a|\le 1$ with $S_a$ constant. 
The vertices of this polygon are precisely ($\pm$) free theory and the yellow point. These are the only points that touch the boundary of the slate. (No other points could touch it since the slate is a convex space.)

\item As we move along the boundary we observe that all S-matrices saturate the unitarity condition at all values of energy \textit{except} for the yellow point discussed above. Unitarity saturation was previously a puzzle in the S-matrix bootstrap approach but as already anticipated in the introduction, it has a nice simple explanation arising from a vanishing duality gap in convex optimization problems together with analyticity. What is particularly nice is that even the exceptional yellow point can be nicely explained in these terms as discussed in the next section. 

\item Perhaps the most striking and still mysterious feature of the S-matrices on the boundary of the slate is that they
are periodic in $\theta$. The period is plotted in figure \ref{fig_period}. 
It is a feature of the slate boundary but it is \textit{not} a generic feature of the S-matrix space boundary; it is not a property of a generic solution at the boundary of the three-dimensional monolith for example. Still, even there, there is some more refined version of emergent periodicity which we comment on in appendix~\ref{app_gral3D}. 

\item Given the periodic nature of the S-matrices at the boundary of the slate, it is natural to explore its inside by considering ansatze  with a fixed period. This can be done quite easily using Fourier coefficients as explained in appendix~\ref{app_periodiccode}. 
Given a particular period, the allowed region touches the boundary of the slate at the points where the S-matrices have the same period but otherwise describes a smaller region inside (since we are not working with the most general S-matrix). This is how the inside curves in figure~\ref{fig_2dcurveB} are generated. Note that already for the period of the periodic Yang-Baxter solution $\tau=2\pi^2/\text{arccosh}(7/2)\approx10.25$ we can approximate very well the boundary of the slate. Also, since free theory and the yellow point are constant solutions, for any period we choose the allowed region will touch the boundary of the slate at those points. In fact, the polygon described above is the extreme case where the period $\tau\rightarrow0$.

\end{itemize}

Apart from the periodicity in $\theta$, what can we say about the poles and zeros of $S_a(\theta)$, i.e. the possible resonances and virtual states? By a careful study of the S-matrices obtained numerically, we were able to understand their analytic structure. A generic S-matrix along the boundary curve of the slate has two different types of analytic structures which we refer to as \textit{simple} and \textit{fractal}.

\begin{figure}[t]
\centering
\includegraphics[width=\textwidth]{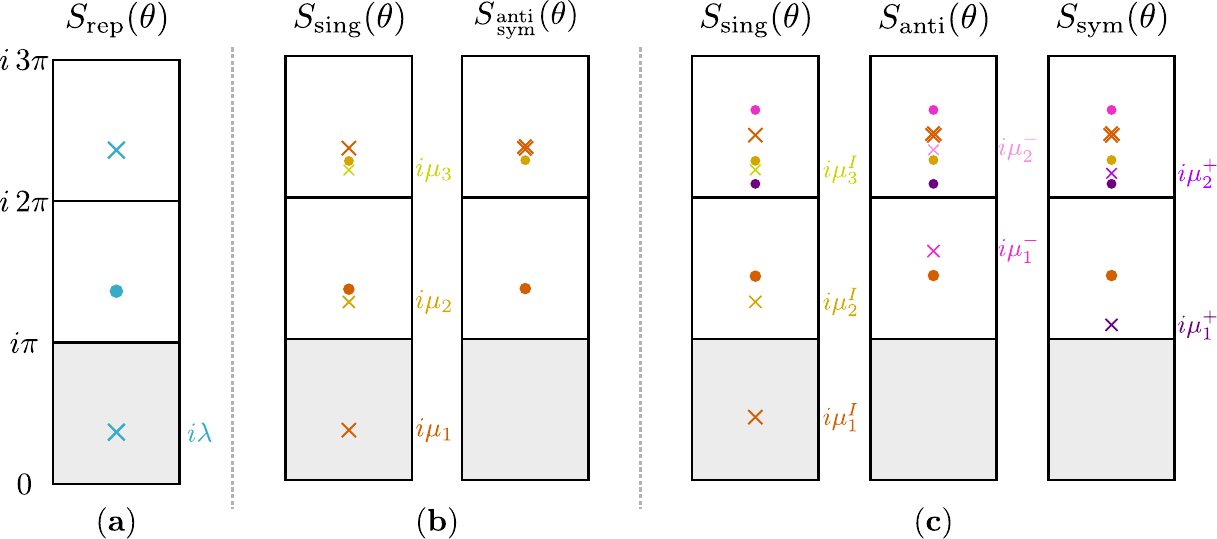}
\caption{Different types of analytic structures appearing in the $\theta$ plane. ({\bf a}) \textit{Simple} structure given by a single ratio of gamma functions $F_{\lambda+\pi}(\theta)$. ({\bf b}) The simplest \textit{fractal} structure with an infinite set of parameters $\mu_i$ labelling the new towers of poles and zeros appearing in higher strips; this is the type of structure present in the green line of figure~\ref{fig_mono} described in appendix~\ref{app_sigma2line}.
({\bf c}) The general \textit{fractal} structure with three (infinite) sets of parameters $(\mu_i^I,\mu_i^-,\mu_i^+)$ according to the representation (sing, anti, sym) on which the first (i.e. closest to the physical strip) zero/pole appears.}
\la{fig_fractal}
\end{figure}

The \textit{simple} structures are the building blocks of the $O(N)$ S-matrices studied here. Starting from an initial pole or zero, we can recover all the poles and zeros in higher sheets from crossing and unitarity as explained in the appendix \ref{app_analytic}. This structure is encoded in a particular ratio of gamma functions we called $F_a(\theta)$ shown in figure~\ref{fig_fractal}~({\bf a}) and which we rewrite here for convenience
\beq
F_a(\theta)\equiv \frac{\Gamma \left(\frac{a+i \theta }{2 \pi }\right) \Gamma \left(\frac{a-i \theta +\pi }{2 \pi }\right)}{\Gamma \left(\frac{a-i \theta }{2 \pi }\right) \Gamma \left(\frac{a+i \theta +\pi }{2 \pi }\right)} \,.
\eeq
The integrable solutions can be conveniently written in terms of these simple structures, see above. 
Note that each solution has a single parameter ($\lambda_\text{GN}$ for NLSM and $\nu$ for pYB) and that the infinite product in \eqref{S_pYB} takes care of the periodicity in the real $\theta$ direction.

On the other hand, the \textit{fractal} structures require the inclusion of infinite parameters labeling the new structures emerging as we move to higher sheets. The simplest of these structures appeared in the analytic solution found in \cite{Cordova:2018uop} which depends on an infinite number of parameters $\mu_i$ (see figure~\ref{fig_fractal}~({\bf b})). The general \textit{fractal} structure appearing in the S-matrices has new towers in each representation, leading to three infinite sets of parameters (one per representation) as shown in figure~\ref{fig_fractal}~({\bf c}). 

To take into account the periodicity, the S-matrices are given by a collection of fractal or simple structures appearing either at multiples of the period $\text{Re}(\theta)=n \tau$ or at $\text{Re}(\theta)=(n+\tfrac{1}{2}) \tau$, with $n\in\mathbb Z$. It is a beautiful story how these intricate structures move in the complex $\theta$ plane interpolating between the simpler integrable solutions. In Appendix~\ref{analyticPlane} we explain in detail how this interpolation occurs.

\section{Dual Problem}
\label{analyticdual}

The space of 2-particle S-matrices allowed by the unitarity, crossing and symmetry constraints is convex. In such space we maximize a linear functional. Since the space is convex, there are no local maxima other than a global maximum found at the boundary of the space allowing us to map out such boundary. As we describe in Appendix \ref{app_dp} (see also \cite{boyd2004convex,mosek}) for the case of a general convex maximization problem with a finite number of variables, it is useful to define a so--called dual minimization problem. By taking a continuum limit we can obtain the dual problem we are interested in. Equivalently, as we describe in this section, there is also a simple and straightforward way to derive the same dual problem directly in the infinite dimensional case used to find the S-matrices. In this section we introduce such derivation as well as important consequences that can be derived from it. We start, as before, by defining a functional $\mathcal{F}$ on the space of S-matrices $S_a(\theta)$ that are analytic on the physical strip $0\le \Im(\theta)\le \pi$ and respect crossing symmetry $S_a(i\pi-\theta) =\sum_b C_{ab} S_b(\theta)$ and unitarity $|S_a(\sigma\in\mathbb{R})|\le 1$:
\beq
 \mathcal{F}[S_a] = \sum_a
 n_a\, \Re \[S_a\(\frac{i\pi}{2}\)\] \label{dp1}
\eeq
The sum is over the three representations (singlet, antisymmetric and symmetric traceless) and
 we write the sum explicitly since we do not always have repeated indices.

For simplicity we chose to evaluate the functions at the (unphysical) crossing symmetric point $\theta=\frac{i\pi}{2}$ and therefore we should take $n_a=\sum_b C_{ba} n_b$ without loss of generality since the anti-crossing symmetric part cancels. For a given $n_a$ we can maximize the functional numerically as we already discussed obtaining the curve displayed in figure~\ref{fig_2dcurve}. In particular we obtain a point where the normal to the curve is parallel to $n_a$ (after projecting $n_a$ onto the $\sigma_{1,2}$ plane). Since the curve has kinks, several values of $n_a$ can lead to the same point at the boundary of this two dimensional section we called the $O(N)$ slate. 
We find kinks at the free theory and the integrable $O(N)$ non-linear sigma model. 

Now let us derive the dual minimization problem and its main properties. 
Consider a set of three functions $K_a(\theta)$ analytic on the physical strip except for a pole at $\theta=\frac{i\pi}{2}$ with residue $\mbox{Res}[K_a,\frac{i\pi}{2}]=n_a$. We can then rewrite the functional to maximize as a contour integral along a contour\footnote{The small vertical segments at $\pm \infty$ can be safely dropped {since we require $K(\theta)$ to go to zero there}.} $\calc=(-\infty,\infty)\cup(i\pi+\infty,i\pi-\infty$):
\beqa
\mathcal{F} &=& \sum_an_a\, \Re \[S_a\(\frac{i\pi}{2}\)\] = \sum_a \Re\left[\frac{1}{2\pi i}\oint_{\calc} K_a(\theta)S_a(\theta)d\theta\right]  \\
  &=& \sum_a \Re \left[\frac{1}{2\pi i}\int_{-\infty}^{+\infty}\left(K_a(\sigma)S_a(\sigma)d\sigma-K_a(i\pi-\sigma)S_a(i\pi-\sigma)\right)d\sigma\right]   \label{dp2}
\eeqa
By crossing symmetry we have $ S_a(i\pi-\sigma) = C_{ab} S_b(\sigma) $. 
If we further impose
\beq
 K_a(i\pi-\sigma) = -C_{ba}\, K_b(\sigma)  \label{dp4}\,,
\eeq
namely that $K_a$ obeys anti-crossing with the transpose matrix $C^\intercal$, then both integrals have the same value (since $C^2=1$) and we can write the functional as an integral over the real axis where $S_a$ satisfies the unitarity constraint $|S_a(\sigma)|\le 1$. Thus we get the bound
\beq
 \mathcal{F} = \sum_a \frac{1}{\pi}\int_{-\infty}^{+\infty} \Im\left[K_a(\sigma)S_a(\sigma)\right] d\sigma 
 \le \sum_a \frac{1}{\pi}\int_{-\infty}^{+\infty} \left|K_a(\sigma)S_a(\sigma)\right| d\sigma
 \le \frac{1}{\pi}\int_{-\infty}^{+\infty}  \sum_a \left|K_a(\sigma)\right| d\sigma   \label{dp5}
\eeq 
where the right hand side is the definition of the dual functional on the space of $K_a$. Thus, we obtain
\beq
\boxed{
\max_{\{S_a\}}\left[\mathcal{F}= \sum_a
 n_a\, \Re \[S_a\(\frac{i\pi}{2}\)\]\right] \le \min_{\{K_a\}}\left[\mathcal{F}_d \equiv  \frac{1}{\pi}\int_{-\infty}^{+\infty}  \sum_a \left|K_a(\sigma)\right| d\sigma \right]  \label{dp6}}
\eeq
where the maximum is over all functions $S_a(\theta)$ analytic on the physical strip and obeying crossing and the unitarity constraint and the minimum is over all functions $K_a(\theta)$ analytic on the physical strip except for a pole at $\theta=\frac{i\pi}{2}$ with residue $\mbox{Res}[K_a,\frac{i\pi}{2}]=n_a$ and obeying anticrossing with $C^\intercal$. To be more precise, we can add the condition that $S_a$ are bounded analytic functions (from the unitarity constraint) whereas $K_a$ are only required to be such that $\mathcal{F}_d$ is finite, namely $\int_{-\infty}^{+\infty} |K_a(\sigma)| d\sigma < \infty$. The minimization problem is also a convex optimization problem known as the dual of the original or primal problem. 
The difference between the minimum of the dual problem and the maximum of the primal problem is called the duality gap. If the primal problem is convex and the dual strictly feasible\footnote{Strict feasibility of the dual problem is sometimes called Slater's condition. It means that there is a point in the interior of the dual cone that satisfies the linear constraints. In this case it means that there is at least a set of functions $K_a$ that satisfy all conditions. In the next section we give the example $K_a=\frac{in_a}{\cosh\theta}$.} (as is the case here), the duality gap vanishes \cite{boyd2004convex} implying that the inequalities in eq.(\ref{dp5}) are saturated. Therefore we must have for every $\sigma\in \mathbb{R}$ and every representation $a$:
\beqa
 |K_a(\sigma) S_a(\sigma)| &=& |K_a(\sigma)| \ \ \ \Rightarrow \ \ \ |S_a(\sigma)|=1 \ \ \mbox{or} \ \ K_a(\sigma)=0  \label{dp7}\\
 \Re\left[K_a(\sigma) S_a(\sigma)\right] &=& 0     \label{dp8} \\
 \Im\left[K_a(\sigma) S_a(\sigma)\right] &\ge& 0     \label{dp9}
\eeqa 
 Since $K_a$ is analytic, if it vanishes on a segment of the real axis, it will vanish everywhere in the physical strip. If that is not the case, it implies that $|S_a(\sigma)|=1$, namely unitarity is saturated everywhere at the maximum of the functional. It is in principle possible that $K_a(\sigma)=0$ at isolated points but, assuming continuity of $|S_a(\sigma)|$ we will still have $|S_a(\sigma)|=1$ on the real axis (physical line). Furthermore, assuming that $K_a(\sigma)\neq 0$, the only way to satisfy the other two conditions is that
\beq
S_a = i\frac{
K^*_a(\sigma)}{|K_a(\sigma)|}   \label{dp10} 
\eeq
 providing a simple way to determine the S-matrix once the dual problem is solved, and also making evident it saturates unitarity. Before continuing let us summarize some simple but useful properties of the dual problem:
\begin{itemize}
 \item In the dual problem there are no inequality constraints for $K_a$ so finding the minimum is generically an easier task. For the numerics in this paper we used the discretized version described in appendix \ref{app_dp} or the Fourier decomposition parametrization in appendix \ref{app_periodiccode}. 
 \item Taking $K_a$ within a subset of all analytic functions (except for the pole at $\theta_*$) one can put upper bounds that will always be larger or equal than the best upper bound. This can be done sometimes analytically and is complementary to taking $S_a$ on a subset which will give a value below the best upper bound. In this way one can bracket the optimal bound  as shown in figure \ref{fig_2dcurveB}. 
 \item If both extremal functions $S_a$ and $K_a$ are obtained analytically, a zero duality gap is an analytic proof that such $S_a$ indeed maximize the given functional.  
 \item Using the previous point, if one can show analytically that a given $S_a$ maximizes different functionals, then one has a proof that the convex set of allowed $S_a$ has a vertex at that point (at least in the considered subspace). 
\end{itemize}
\subsection*{Applications and Generalizations}

 We can illustrate the last bullet point with the simple example of free theory where $S_a(\theta)=1$. In particular the curve in figure~\ref{fig_2dcurve} has a kink at the free theory as we can now derive analytically. The value of the functional (\ref{dp1}) is just
\beq
 \mathcal{F} = \sum_a n_a\,.  \label{dp11}
\eeq
 For $K_a$ we can take the simple ansatz
\beq
 K_a = \frac{i n_a}{\cosh\theta} \,,   \label{dp12}
\eeq
which has a simple pole at ${i\pi}/{2}$ with residue $n_a$ (all other poles are outside the physical strip). Using that $\cosh(\sigma)>0$ the dual functional $\mathcal F_d$
can be easily evaluated to get
\beq
  \mathcal{F}_d=\frac{1}{\pi}\int_{-\infty}^{+\infty}  \sum_a \left|K_a(\sigma)\right| d\sigma =\sum_a |n_a| \, \frac{1}{\pi}\int_{-\infty}^{+\infty} \frac{1}{\cosh(\sigma)} d\sigma  =\sum_a |n_a|\,.    \label{dp13}
\eeq
Indeed, this is the simplest example of \eqref{dp6} since
\beq
   \mathcal{F} = \sum_a n_a \le \mathcal{F}_d=\sum_a |n_a|\,.   \label{dp14}
\eeq
 The inequality is saturated when $n_a
 \ge 0$. 
Furthermore, to satisfy the anti-crossing condition \eqref{dp4} we need $n_a=C_{ba} n_b$. Then, up to an overall normalization, $n_a$ takes the form
\beq
(n_\text{sing},n_\text{anti},n_\text{sym}) = \(0,\,\half,\,\half\) + \alpha \(\frac{1}{2N},-\frac{1}{4},\frac{N-2}{4N}\), \ \ \ \ \ 0\le\alpha\le 2.     \label{dp15}
\eeq
 For all the $n_a$ above, the functional is maximized by the free theory showing that the free theory is indeed at a kink of the boundary curve as seen in fig. \ref{fig_2dcurve}. 
 
In fact the test functions $K_a = {i n_a}/{\cosh\theta}$ can be used to put  upper bounds for all directions in the $\sigma_{1,2}$ plane. Indeed, consider the following maximization problem
\beq
 \max_{\{S_a\}}[t], \ \ \mbox{such that} \ \ \Re\[S_a\(\frac{i\pi}{2}\)\]=t\, v_a, \ \ \ \sum_a v_a^2=1.       \label{dp16}
\eeq
Finding the maximum of $t$ and replacing in $S_a(\frac{i\pi}{2})=t\, v_a$ determines a point on the boundary curve of figure~\ref{fig_2dcurve} in the direction $v_a$ (projected on the plane $\sigma_{1,2}$). Namely we find a point in a given direction rather than a point with a given normal as was the case when fixing $n_a$ as discussed earlier, see figure \ref{3functionals}. We can write a Lagrangian using Lagrange multipliers $\mu_a$:
\beq
 L = t + \sum_a \mu_a \left\{\Re \[S_a\(\frac{i\pi}{2}\)-t v_a\]\right\} = t \(1-\sum_a \mu_a v_a\) + \sum_a \Re\left[\frac{1}{2\pi i}\oint_{\calc} K_a(\theta)S_a(\theta)d\theta\right]      \label{dp17}
\eeq
where we take $K_a$ as before with $\mbox{Res}[K_a,\frac{i\pi}{2}]=\mu_a$. Maximizing $L$ over the space of $S_a$ satisfying the constraint is the same as maximizing $t$ since then $L=t$ independently of the value if $\mu_a$. If we choose $\mu_a$ such that
\beq
 \sum_a \mu_a v_a =1      \label{dp18}
\eeq
 then we have 
\beq
 L = \sum_a \Re\left[\frac{1}{2\pi i}\oint_{\calc} K_a(\theta)S_a(\theta)d\theta\right] \le \frac{1}{\pi}\int_{-\infty}^{+\infty}  \sum_a \left|K_a(\sigma)\right| d\sigma      \label{dp19}
\eeq
where we used the same bound derived before in (\ref{dp5}). We learn that 
\beq
\max_{\{S_a\}}[t] \le \min_{\{K_a\}}\left[  \frac{1}{\pi}\int_{-\infty}^{+\infty}  \sum_a \left|K_a(\sigma)\right| d\sigma \right]      \label{dp20}
\eeq
where the maximum is over all $S_a$ satisfying the extra constraints in (\ref{dp16}) and the minimum over all $K_a$ with residue $\mu_a$ satisfying (\ref{dp18}). This minimization problem can be used numerically to calculate the boundary curve in fig. \ref{fig_2dcurve}. 

If we just consider the simple functions $K_a = \frac{i n_a}{\cosh\theta}$ we find an exterior curve determined by the minimization problem:
\beq
 \min_{n_a} \sum_a |n_a| , \ \ n_a v_a=1,\ n_a=\sum_b C_{ba} n_b        \label{dp21}
\eeq
In each region where $n_a$ have definite signs, the function to minimize is linear and therefore it is minimized at the boundary of the region, namely where one $n_a$ vanishes. By enumerating the different possibilities one finds the bound given by the enveloping polygon in figure \ref{fig_2dcurveB}, whose vertices are
\beq
\(\sigma_2,\sigma_1\)_\text{vertices} = \left\{ \(1,0\), \(0,\half\), \(-1,\frac{1}{N}\), \(-1,0\), \(0,\half\), \(1,-\frac{1}{N}\)     \right\}   \label{dp22}
\eeq
We now consider the possibility of $S_a$ not saturating unitarity. As already discussed this can happen only if $K_a$ is identically zero for some representation $a$. In particular the corresponding residue $n_a$ has to vanish as well. Taking $n_\text{sing}=0$ or $n_\text{anti}=0$ leads to the free theory. 
For the remaining case $n_\text{sym}=0$ we get something more interesting. Using crossing we can determine up to an overall constant
\beq
 (n_\text{sing},n_\text{anti},n_\text{sym})= (1,1-N,0)\,.     \label{dp23}
\eeq 
 If we take again the simple functions $K_a = \frac{i n_a}{\cosh\theta}$ then from (\ref{dp10}) we have $S_\text{sing}=1$ and $S_\text{anti}=-1$ implying that $S_a$ are constant and
\beq
 \mathcal{F} = S_\text{sing} + (1-N) S_\text{anti} = N \le |1| + |N-1| = N     \label{dp24}
\eeq 
Since the inequality is saturated we learn that the constant functions indeed maximize this functional. On the other hand using crossing we obtain $S_\text{sym}=-\frac{N-2}{N+2}$ which does not saturate unitarity ($|S_\text{sym}|<1$) consistently with $K_\text{sym}=0$. This is precisely the \textit{yellow point}  discussed above.
 
 It is also interesting to consider the case where we evaluate the S-matrix at a different interior point. Using crossing symmetry we can define such functional as
\beq
 \mathcal{F} =  \half \sum_a
  \left(n_a \Re \[S_a(\theta_* )\] + \sum_b
   n_b C_{ba} \Re \[S_a(i \pi-\theta_* )\] \right)     \label{dp25}
\eeq
Due to crossing symmetry both terms are equal so that $\mathcal F =  \sum_a
n_a \Re \[S_a(\theta_* )\]$. Using the previous reasoning, we choose $K_a$ to have poles at $\theta_*$ and $i\pi-\theta_*$ with residues $\mbox{Res}[K_a,\theta_*]=n_a$ and  $\mbox{Res}[K_a,i\pi-\theta_*]=\sum_b C_{ba}n_b$.
Under those conditions (\ref{dp5}) is still valid and can be used to find $S_a$, put bounds, etc. In the same way, (\ref{dp20}) is also valid. 

 Going back to the case where we evaluate the functional at the crossing symmetric point $\theta=i\frac{\pi}{2}$, using the dual problem we can bracket the optimal bound as seen in figure~\ref{fig_2dcurveB}. In that figure, the black curve is the optimal bound.
To obtain the interior curves, we take the S-matrices as  periodic functions along the real axis:
\beq
 S_a(\theta+\tau) = S_a(\theta), \ \ \ \tau\in\mathbb{R}
\eeq
 Maximizing the functional in this set of functions we obtain a maximum that is always smaller or equal than the optimal bound. In that way we draw the interior curves. In particular if we consider constant S-matrices we find the interior polygon contained in all other curves. Appendix \ref{OriginalD} shows a simple numerical implementation of this primal problem, ready to be copy/pasted into \verb"Mathematica".
For the exterior curves we consider functions $K_a$ of the form
\beq
 K_a(\theta) = \frac{ik_a(\theta)}{\cosh\theta}, \ \ \ k_a(\theta+\tau)=k_a(\theta), \ \ \ \ k_a\(\frac{i\pi}{2}\)=n_a  
\eeq
which parameterize a subset of all possible functions $K_a$. Notice however that $K_a$ itself is not periodic, otherwise it would have had infinite number of poles on the line $\Im(\theta)=\frac{\pi}{2}$ instead of just one as required. Numerically minimizing the dual function we find the exterior curves. In the particular case of constant $k_a$ we obtain the exterior polygon that contains all other curves and that was derived in more detail in the previous subsection. Appendix \ref{DualD} contains  a simple numerical implementation of this dual problem, ready to be copy/pasted into \verb"Mathematica".

\vspace{0.5cm}

 Summarizing, in this section we derived the dual problem that allows us to explain why the maximum generically saturates unitarity on the physical line, also allows us to bracket the optimal bound, an important point since results are usually numeric, and finally provides a procedure to check when a given analytic function $S_a$ maximizes a given functional.


\section{Discussion}

 In this paper we considered the scattering matrices of massive quantum field theories with no bound states and a global $O(N)$ symmetry in two spacetime dimensions. In particular we explored the space of two-to-two S-matrices of particles of mass $m$ transforming in the vector representation as restricted by the general conditions of unitarity, crossing, analyticity and $O(N)$ symmetry. Such space is an infinite dimensional convex space parameterized by three analytic functions $S_a
 (s)$ of the Mandelstam variable $s$. The index $a$
  indicates the $O(N)$ representation to which the initial two particle state belongs: singlet, antisymmetric or symmetric traceless.
  A simple picture of that space can be obtained by finding all the allowed values of the functions $S_a(s_*)$
  at an unphysical point $0<s_*<4m^2$. In this way we obtain a three-dimensional convex subspace which we dub as the $O(N)$ \textit{monolith} that can be plotted using numerical methods. A beautiful picture emerges and at the boundary of this space we identify vertices that correspond to known theories (free theory and the integrable $O(N)$ non-linear sigma model). Another interesting theory appears at a point we call a pre-vertex, an intersection of two edges but with no curvature singularity. Finally there is an interesting point corresponding to a constant solution that does not saturate unitarity in one of the channels. This is an exceptional case since at all other boundary points the S-matrices obtained saturate unitarity. Although the results are numeric for several points we find analytic expressions for the S-matrix including a line that connects two integrable points. In the particular case of the crossing symmetric point $s_*=2m^2$, the crossing anti-symmetric linear combination vanishes and the space of allowed values is two dimensional, and now dubbed as the $O(N)$ \textit{slate}. Again we obtain an interesting boundary contour with vertices at the free theory and $O(N)$ non-linear sigma model. A curious property of this case is that the S-matrices at the boundary curve of the $O(N)$ \textit{slate} are periodic in the rapidity. 
 
 A simple way to find the boundary of the allowed space is to maximize a linear functional in the convex space since the maximum is always at the boundary. In general convex maximization problems the so call dual problem plays an important role. The same happens in this case. Indeed we find that the dual problem consists of minimizing a functional over the space of analytic functions with a pole at $s_*$ (the point where we evaluate the S-matrix). The main property of the dual problem is that the minimum of the dual functional equals the maximum of the original one for convex problems such as this one. This allows for some important numerical an analytical results that can be obtained from the dual problem. Numerically, the dual problem has no inequality constraints so it is easier to solve. Also any test function provides a strict upper bound that approaches the boundary of the space from outside as a better ansatz for the functions are found. Additionally, it can be shown that the S-matrices resulting from this problem always saturate unitarity except in the case where the corresponding dual function identically vanishes. This is an exceptional case and corresponds to the constant solution previously discussed. Finally, if the dual functions are found analytically this provides an analytical proof that certain given $S$-matrices maximize the original functional. In fact this can be used to show that the space has vertices by showing that different functionals are maximized by the same $S$-matrices.  
 
  In summary, we found a rich structure in the allowed space of S-matrices for two dimensional massive theories with particles in the vector representation of $O(N)$ by using convex maximization and in particular its convex dual minimization problem. At the boundary of the allowed space special geometric points such as vertices (and pre-vertices as defined above) were found to correspond to integrable models. Although the dual minimization problem implies that unitarity is saturated as it should be for  integrable models, the reason that such models appear at geometrically distinguished points (e.g. vertices) is not clear. In particular it will be nice to understand if the dual functions play a role in the integrable structure associated with those models.

Finally, in higher dimensions similar unitarity saturation was also observed \cite{Paulos:2017fhb,Guerrieri:2018uew}.  It would be very interesting to develop the higher dimensional dual problem which should explain this saturation, see also \cite{Caprini_1980,Lopez:1975ca,Lukaszuk:1967zz,Bonnier:1968zz,Bonnier:1975xu}. At the same time, it is known that unitarity can not be  saturated at all energies and spins in higher dimensions \cite{NoPPNoGood}. It would be fascinating to resolve this tension and find a sharp rigorous dual problem in higher dimensions.

\section{Acknowledgments}
We thank Carlos Bercini, Frank Coronado, Matheus Fabri, Andrea Guerrieri, Matt Headrick, Alexandre Homrich, Joao Penedones, Balt van Rees, Jon Toledo, Sasha Zamolodchikov for very useful discussions. 
M.K. is grateful 
the DOE that supported in part this work through grants DE-SC0007884 and DE-SC0019202 as well to the Keck Foundation that also provided partial support for this work. The work of M.K. also benefited from the 2019 Pollica summer workshop, which was supported in part by the Simons Foundation (Simons Collaboration on the Non-perturbative Bootstrap) and in part by the INFN. Research at the Perimeter Institute is supported in part by the Government of Canada through NSERC and by the Province of Ontario through MRI. 
This work was additionally supported by a grant from the Simons Foundation (PV: \#488661) and FAPESP grants 2016/01343-7 and 2017/03303-1.

\appendix

\section{Notation}\la{notation}
Crossing matrix 
\beq
C_{ab}=\left(
\begin{array}{ccc}
\frac{1}{N}&-\frac{N}{2}+\frac{1}{2}\,\,&\frac{N}{2}+\frac{1}{2}-\frac{1}{N}\\
-\frac{1}{N}&\frac{1}{2}&\frac{1}{2}+\frac{1}{N}\\
\frac{1}{N}&\frac{1}{2}&\frac{1}{2}-\frac{1}{N}
\end{array}
\right) \,.
\la{Cmatrix}
\eeq
where $a,b=\text{sing},\text{anti},\text{sym}$.

Two different decompositions of the two-to-two S-matrices are:
\beqa
\mathbb{S}(\theta)&=&\sigma_1(\theta)\mathbb{K}+\sigma_2(\theta)\mathbb{I}+\sigma_3(\theta)\mathbb{P}\la{sigma_decomposition}\\
&=&S_{\text{sing}}(\theta)\mathbb{P}_{\text{sing}}+S_{\text{anti}}(\theta)
\mathbb{P}_{\text{anti}}+S_{\text{sym}}(\theta)\mathbb{P}_{\text{sym}} \la{rep_decomposition}\,,
\eeqa
where $\mathbb{K}_{ij}^{kl} = \delta_{ij}\delta^{kl},\,\mathbb{I}_{ij}^{kl} = \delta_{i}^l\delta_j^{k},\, \mathbb{P}_{ij}^{kl} = \delta_{i}^{k}\delta_j^{l}$. The bases are related by the trivial map:
\beqa
S_{\text{sym}}=\sigma_2+\sigma_3\,,\qquad
S_{\text{anti}}=\sigma_2-\sigma_3\,,\qquad
S_{\text{sing}}=N\sigma_1+\sigma_2 +\sigma_3 \,.\la{rep123}
\eeqa

\section{The Primal-dual Quadratic Conic Optimization}
\label{app_dp}
In this appendix, we review the standard primal-dual conic optimization problem and its relation to the S-matrix bootstrap we studied in the main text. In particular we consider the discretized version of the dual problem described in section \ref{analyticdual}. See references \cite{boyd2004convex,mosek} for more details on convex optimization.\footnote{For parts of the optimization, we used CVX, a package for specifying and solving convex programs \cite{cvx,gb08}.}

The standard conic optimization problem is given by
\begin{equation}\label{primal}
\begin{aligned}
\text{minimize} \;\; & c^T x\\
\text{subject to} \;\; & Ax=b\\
& x \in \mathcal{K}
\end{aligned}
\end{equation}
where $\mathcal{K}$ is a convex cone. One can then write down the Lagrangian
\begin{equation}
L(x,\lambda,\nu)=c^T x+\nu^T (Ax-b)-\lambda^T x
\end{equation}
where $\nu$ is the Lagrange multiplier of the linear constraint and $\lambda \in \mathcal{K^*}$ is the dual cone satisfying
\begin{equation}
\lambda^T x \ge 0,\; \forall x\in\mathcal{K},\; \lambda\in\mathcal{K^*}.
\end{equation}
The dual function is defined as
\begin{equation}
g(\lambda,\nu)=\inf_{x}{L(x,\lambda,\nu)}=\begin{cases}
-b^T \nu,\; c+A^T\nu-\lambda=0\\
-\infty ,\; \text{otherwise}
\end{cases}
\end{equation}
and the dual problem is
\begin{equation}\label{dual}
\begin{aligned}
\text{maximize} \;\; & -b^T \nu\\
\text{subject to} \;\; & c+A^T\nu-\lambda=0\\
& \lambda \in \mathcal{K}^*
\end{aligned}
\end{equation}
For any feasible point of the primal and dual problem $(\tilde{x}, \tilde{\lambda},\tilde{\nu})$ one has
\begin{equation}\label{dualitygap}
-b^T \tilde{\nu}=-\tilde{x}^T A^T\tilde{\nu}=c^T \tilde{x}-\tilde{\lambda}^T \tilde{x}\le c^T \tilde{x}
\end{equation}
where we used the linear constraints of the primal and dual problem for the first and second equality. The difference between the maximum of the dual function $\tilde{g}$ and the minimum of the primal function $\tilde{f}$ is $\tilde{\lambda}^T\tilde{x}$ which is the so-called duality gap.

In the S-matrix bootstrap, we discretize the S-matices by its values on the physical line $S_a(\sigma_i),i=1,...,M$. The unitarity constraints are 
\begin{equation}
\Re S_A^2+\Im S_A^2\le1,\; A=(a,i).
\end{equation}
It is convenient to consider the rotated quadratic cones instead:
\begin{equation}
\Re S_A^2+\Im S_A^2\le 2 u_A v_A
\end{equation}
with the trivial linear constraints
\begin{equation}\label{trivialuv}
u_A=\frac{1}{\sqrt{2}},\; v_A=\frac{1}{\sqrt{2}},\; \forall A.
\end{equation}
The real and imaginary parts are related by
\begin{equation}
\Im S_A=\mathbb{K}_{AB}\Re S_A
\end{equation}
which is the discrete version of the dispersion relation together with crossing constraint. (See definition in \cite{He:2018uxa}.) Therefore we can write our bootstrap problem in the standard quadratic conic optimization language with the following identifications:
\begin{equation}\label{primalid}
x=\begin{pmatrix}
\Re S\\
\Im S\\
u\\
v\\
\end{pmatrix},\; 
A=\begin{pmatrix}
\mathbb{K}&-\mathbb{I}& 0 & 0\\
0 & 0 & \mathbb{I} & 0\\
0 & 0 & 0 & \mathbb{I}\\
\end{pmatrix},\;
b=\begin{pmatrix}
0\\
\frac{1}{\sqrt{2}}\\
\frac{1}{\sqrt{2}}\\
\end{pmatrix}.
\end{equation}
The elements in $x$ and $b$ should be understood as $3M$-dimensional column vectors and the elements in $A$ are $3M\times3M$-dimensional matrices. For any given  maximization in the 2d and 3d plots, the functional can be written as
\begin{equation}\label{primalfun}
\mathcal{F}=\sum_A w_A \Re S_A,
\end{equation}
and hence
\begin{equation}
c=\begin{pmatrix}
-w\\
0\\
0\\
0\\
\end{pmatrix}.
\end{equation}
With these identifications, we can consider the dual variables
\begin{equation}
\lambda=\begin{pmatrix}
\lambda_{1}\\
\lambda_{2}\\
\lambda_{3}\\
\lambda_{4}\\
\end{pmatrix},\;
\nu=\begin{pmatrix}
\nu_{1}\\
\nu_{2}\\
\nu_{3}\\
\end{pmatrix}
\end{equation}
of the dual problem \eqref{dual}. The dual cones are given by:
\begin{equation}
\lambda_{1A}^2+\lambda_{2A}^2\le 2\lambda_{3A}\lambda_{4A},\;\lambda_{3A}>0,\;\lambda_{4A}>0,\;\forall A
\end{equation}
where we used that these quadratic cones are self-dual. With the explicit expressions of \eqref{primalid}, the dual linear constraint and the dual functional become the following:
\begin{equation}
\begin{aligned}
\mathcal{F}_d&=-\sum_A\frac{1}{\sqrt{2}}(\nu_{2A}+\nu_{3A})\\
w_A&=\mathbb{K}^T_{AB}\nu_{1B}-\lambda_{1A}\\
-\nu_{1A}&=\lambda_{2A}\\
\nu_{2A}&=\lambda_{3A}\\
\nu_{3A}&=\lambda_{4A}\\
\end{aligned}
\end{equation}
which reduces to
\begin{eqnarray}
\mathcal{F}_d&=&-\sum_A\frac{1}{\sqrt{2}}(\lambda_{3A}+\lambda_{4A})\label{dualfun}\\
-w_A&=&\mathbb{K}^T_{AB}\lambda_{2B}+\lambda_{1A}.\label{poleconstr}
\end{eqnarray}

From \eqref{dualitygap} we see that in the primal-dual problem, the duality gap is closed when we have
\begin{equation}
\tilde{\lambda}^T \tilde{x}=0,
\end{equation}
i.e. $\tilde{\lambda}$ and $\tilde{x}$ are orthogonal. It is easy to see that this happens iff $(\tilde{\lambda}_{1A},\tilde{\lambda}_{2A},\tilde{\lambda}_{3A},\tilde{\lambda}_{4A})$ and $(-\Re \tilde{S}_{A},-\Im \tilde{S}_{A}, \tilde{v}_{A}, \tilde{u}_{A})$ are parallel. 
Let us thus write
\begin{equation}\label{parallel}
(\tilde{\lambda}_{1A},\tilde{\lambda}_{2A},\tilde{\lambda}_{3A},\tilde{\lambda}_{4A})=\tilde{\kappa}_A(-\Re \tilde{S}_{A},-\Im \tilde{S}_{A}, \tilde{v}_{A}, \tilde{u}_{A}),
\end{equation}
and
\begin{equation}\label{dualitygapclose}
\begin{aligned}
\tilde{\lambda}^T \tilde{x}&=\tilde{\lambda}_{3A}\tilde{u}_A+\tilde{\lambda}_{4A}\tilde{v}_{4A}+\tilde{\lambda}_{1A}\Re \tilde{S}_A+\tilde{\lambda}_{2A} \Im \tilde{S}_{A}\\
&=\sum_A \tilde{\kappa}_A(2\tilde{u}_A \tilde{v_A}-\Re \tilde{S}_A^2-\Im \tilde{S}_A^2)\\
&=0.
\end{aligned}
\end{equation}
Since $\tilde{\kappa}_A\ge 0$ and $2\tilde{u}_A \tilde{v_A}\ge\Re \tilde{S}_A^2+\Im \tilde{S}_A^2$, $\forall A$, we see that for the last equality of \eqref{dualitygapclose} to be true we have either $\tilde{\kappa}_A=0$ or $\Re \tilde{S}_A^2+\Im \tilde{S}_A^2= 2\tilde{u}_A \tilde{v}_A=1$, i.e., unitarity saturation for each $A$.
Using \eqref{trivialuv} and \eqref{parallel}, the dual maximization functional \eqref{dualfun} becomes
\begin{equation}
\mathcal{F}_d=-\sum_A \kappa_A.
\end{equation}
To summarize, the optimal value of the primal function \eqref{primalfun} can be obtained by solving the following dual optimization problem
\begin{equation}\label{dualS}
\begin{aligned}
\text{minimize} \;\; & \sum_A \kappa_A\\
\text{subject to} \;\; & -\mathbb{K}^T_{AB}\lambda_{2B}=\lambda_{1A}+w_A.\\
& \lambda_{1A}^2+\lambda_{2A}^2= \kappa_{A}^2
\end{aligned}
\end{equation}

Now let us interpret the dual optimization problem \eqref{dualS}. Following the properties of $\mathbb{K}$, we see that $-\mathbb{K}^T$ gives the dispersion relation with crossing using $-C^T$. Therefore we can identify $\lambda_{2A}$ and $\lambda_{1A}+w_A$ as the real and imaginary parts of an analytic function on the physical line satisfy crossing with $-C^T$
\begin{equation}\label{h}
h_a(i\pi-\theta)=-h_b(\theta)C_{ba}.
\end{equation}
For a maximization in the 2d plot with a fixed normal vector, we have the functional
\begin{equation}
\begin{aligned}
\mathcal{F}=&\sum_a n_a \Re S_a\big(i\frac{\pi}{2}\big)\\
=&\Re\bigg[\frac{1}{2\pi i}\sum_a\oint\frac{i n_a S_a(\theta)}{\cosh(\theta)}\bigg]\\
=&\frac{1}{2\pi}\sum_a\int_{-\infty}^{\infty}d\sigma\frac{n_a\Re S_a(\sigma)+n_a C_{ab} \Re S_b(\sigma)}{\cosh(\sigma)}\\
=&\frac{1}{\pi}\sum_a\int_{-\infty}^{\infty}d\sigma\frac{n_a\Re S_a(\sigma)}{\cosh(\sigma)},
\end{aligned}
\end{equation}
where in the last equation we assume $n_b=n_a C_{ab}$. 
Comparing with \eqref{primalfun}, we see that $w_A, A=a,i$ correspond to the discrete version of the imaginary parts of $\frac{i n_a}{\cosh(\theta)}$ on the physical line which have poles at $i\frac{\pi}{2}$ with crossing symmetric residues. We therefore conclude that the analytic continuation of $\lambda_a$ into the physical strip also include such pole terms with opposite signs to cancel the poles and we have
\begin{equation}
-\lambda_a(\theta)=\frac{i n_a}{\cosh(\theta)}-h_a(\theta).
\end{equation}
The minimization problem \eqref{dualS} reduces to minimizing $\sum_A \kappa_A=\sum_A |\lambda_A|$ where $-\lambda_a$ are analytic functions with poles at $i\pi/2$ and crossing symmetric residues $n_b=n_a C_{ab}$.

We can now make the following identifications
\begin{equation}
\lambda_{1A}\to -\Im K_A,\;\lambda_{2A}\to -\Re K_{A}, \kappa_{A}=|\lambda_A| \to |K_A|.
\end{equation}
Combined with \eqref{parallel}, we see
\begin{equation}
\Re S_A=\frac{\Im K_A}{|K_A|}, \; \Im S_A=\frac{\Re K_A}{|K_A|}.
\end{equation}
This becomes \eqref{dp10} in the continuous limit.

\bigskip

\section{Analytic Properties} \la{app_analytic}

In this appendix we further explain some of the analytic properties of the S-matrices on the boundary of the monolith. A first simple characterization of some of these properties is the value at threshold of the three different channels $S_a(\theta=0)$. Given the generic saturation of unitarity ($S_a(\theta)S_a(-\theta)=1$), the quantity $S_a(\theta=0)$ can take values $\pm1$ leading to the eight possible combinations in table~\ref{tab:colors}, represented in different colors. This is the coloring used in figures~\ref{fig_morph} and~\ref{fig_2dcurve} which highlights some of the geometric aspects of the first one and interesting points of the latter. Apart from geometry, the transition from one color to another indicates changes in the analytic structure of the S-matrices such as collision of zeros and poles at the boundary of the physical strip. Such collisions and further phenomena are explained in detail in the next section for the S-matrices at the boundary of the slate.

\begin{table}[t]
  \centering
\begin{tabularx}{235.2pt}{|cccc|}
\hline
&$S_\text{sing}(\theta=0)$&$S_\text{anti}(\theta=0)$&$S_\text{sym}(\theta=0)$\\
\hline
${\color{black} \blacksquare}$&$-1$&$-1$&$-1$\\
${\color{darkgreen} \blacksquare}$&$-1$&$-1$&$+1$\\
${\color{blue} \blacksquare}$&$-1$&$+1$&$-1$\\
${\color{lightpink} \blacksquare}$&$-1$&$+1$&$+1$\\
${\color{pink} \blacksquare}$&$+1$&$-1$&$-1$\\
${\color{lightblue} \blacksquare}$&$+1$&$-1$&$+1$\\
${\color{lightgreen} \blacksquare}$&$+1$&$+1$&$-1$\\
${\color{gray} \blacksquare}$&$+1$&$+1$&$+1$\\
 \hline
\end{tabularx}
 \caption{Colors assigned to the eight different combinations of $S_a(\theta=0)=\pm1$ which help characterize  the faces and edges of the monolith as in figures~\ref{fig_morph} and~\ref{fig_2dcurve}.}
 \label{tab:colors}
\end{table}


%
%
%


\subsection{The Slate}\la{analyticPlane}

In the following we explain how the analytic structure of the S-matrices changes as we move along the boundary of the $\theta_\star=i\pi/2$ slate of figure~\ref{fig_2dcurve}. The interpolation between the different known S-matrices (free, periodic YB, NLSM, constant) is separated in four regions.
 For simplicity, we present the analysis for the first two strips $0\leq\text{Im}(\theta)\leq2\pi$ in the complex rapidity plane.

{\bf Region I: from Free to periodic YB}\\ 
We start from free theory where the complex $\theta$ plane is devoid of any structure. As soon as we move towards the periodic YB solution on the boundary curve of \ref{fig_2dcurve} we get poles and zeros at $\text{Re}(\theta)=n\tau$ in a fractal structure (see figure~\ref{fig_fractal}(c)). The pair of zero and pole\footnote{Recall that unitarity in the rapidity plane reads: $S_a(\theta)S_a(-\theta)=1$, so that for every zero in $\theta$ there is a pole in $-\theta$.} emerging from $\theta=0$ allows for the change of sign in the antisymmetric channel: $S_\text{anti}(\theta=0)=+1$ in free theory to $S_\text{anti}(\theta=0)=-1$ in this region (that is from grey to light blue in the coloring of table~\ref{tab:colors}). Note that the zeros in the second sheet of $S_\text{sing}$ (in orange) --giving rise to the fractal structure-- are necessary so that there are no poles inside the physical strip. As the period decreases we see as well a simple (see figure~\ref{fig_fractal}(a)) structure at $\text{Re}(\theta)=(n+\tfrac{1}{2})\tau$ starting in the symmetric representation (in purple). The simple green structure at multiples of $i\pi$ does not move in the imaginary $\theta$ direction and is present in most of the curve.
\begin{center}
\includegraphics[width=.75\textwidth]{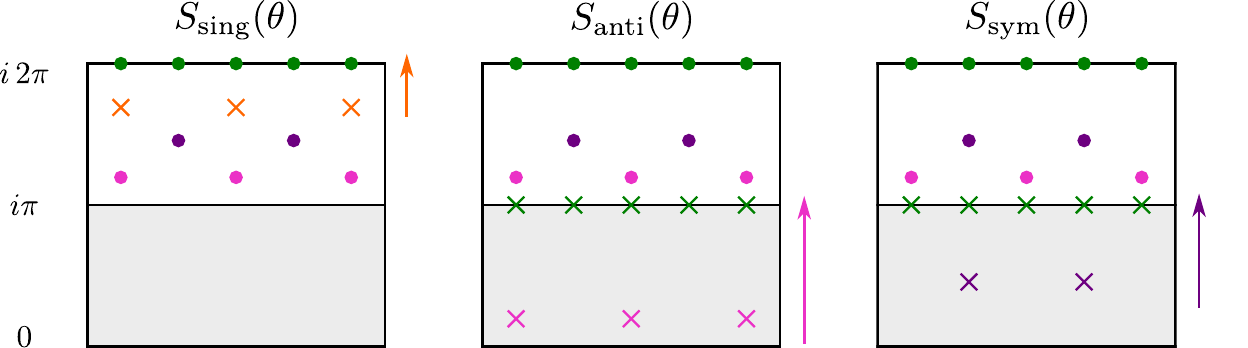}
\end{center}
As we keep moving along the curve, both the fractal and simple structures move into higher sheets indefinitely until disappearing. The period keeps decreasing until it reaches the periodic Yang Baxter value: $\tau=2\pi^2/\nu$. Only the green structure at multiples of $i\pi$ remains, leaving the analytic structure of the periodic Yang Baxter solution \eqref{S_pYB}.
\begin{center}
\includegraphics[width=.75\textwidth]{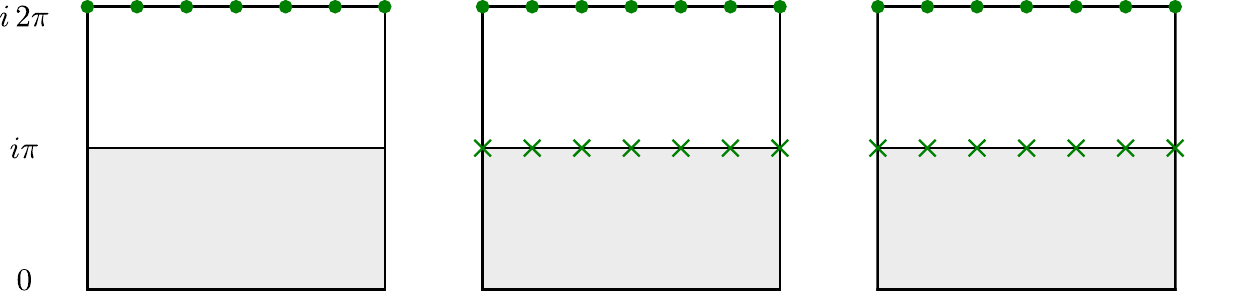}
\end{center}
{\bf Region II: from periodic YB to (-)NLSM}\\
After passing the periodic YB solution, the period again increases as shown in figure~\ref{fig_period}. New structures of fractal type for $\text{Re}(\theta)=(n+\tfrac{1}{2})\tau$ and simple for $\text{Re}(\theta)=n\tau$ come from higher sheets and make their way close to the physical strip.
\begin{center}
\includegraphics[width=.75\textwidth]{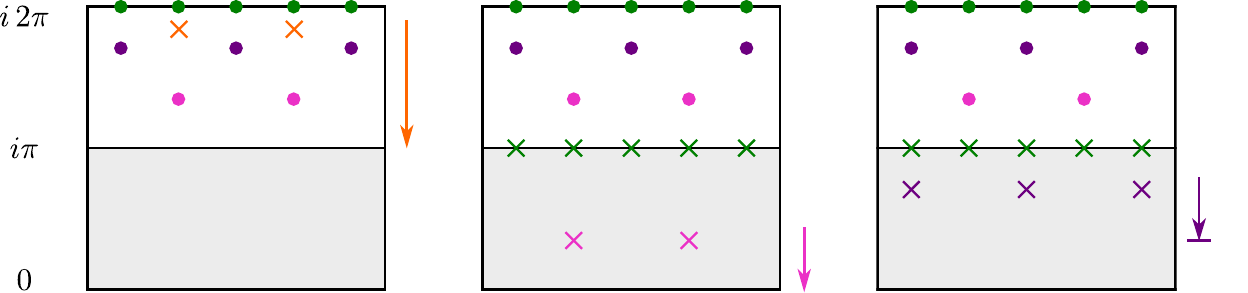}
\end{center}
The structures keep lowering until the zeros in the physical strip of $S_\text{anti}$ (in pink) reach the $\theta\in\mathbb R$ line. In the singlet representation, the fractal structure in orange reaches the line $\theta\in i\pi+\mathbb R$, canceling the dangerous pole at the upper boundary of the physical strip (similar cancellations follow in higher sheets, proving the necessity of the fractal structures). In the symmetric channel, the simple structure (in purple) keeps lowering until it reaches the $\theta\in i \lambda+\mathbb R$ line. In the meantime the period diverges, so that only the central structure remains and we get the NLSM solution  \eqref{S_NLSM}.
\begin{center}
\includegraphics[width=.75\textwidth]{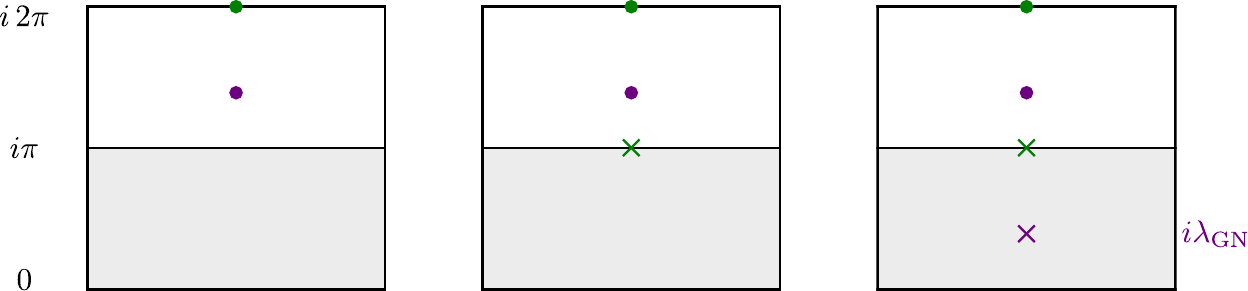}
\end{center}
{\bf Region III: from (-)NLSM to constant solution}\\
As we pass the NLSM the simple structure in $S_\text{sym}$ keeps lowering towards the $\theta\in\mathbb R$ line and at the same time the period decreases. Meanwhile, a new fractal structure emerges from the $\theta\in\mathbb R$ line in the singlet representation, again at $\text{Re}(\theta)=(n+\tfrac{1}{2})\tau$.
\begin{center}
\includegraphics[width=.75\textwidth]{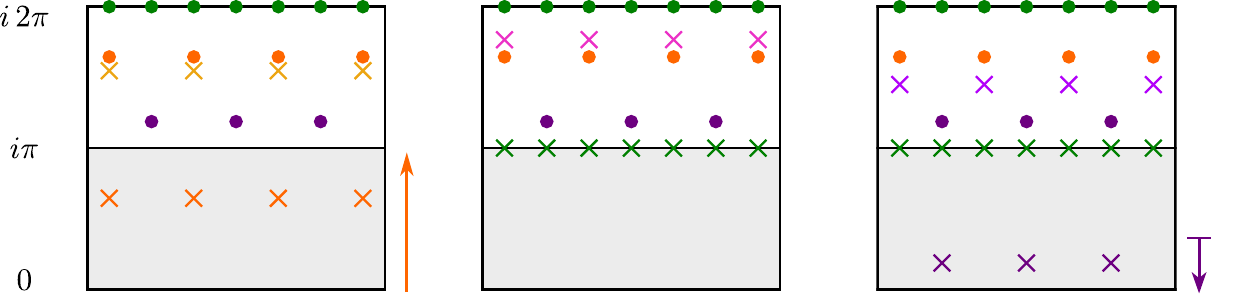}
\end{center}
Now a curious phenomenon occurs: as the simple structure reaches and the fractal one heads to the $\theta\in i\pi+\mathbb R$ line, the period vanishes. This means there is a collision of infinite poles and zeros at $\text{Im}(\theta)=i\pi,2i\pi,...$ and at the real line in the symmetric channel. With this mechanism we reach the constant solution \eqref{constant_soln} where $|S_\text{sym}|<1$!\\
\\
{\bf Region IV: from constant solution to (-)Free}\\
Finally, as the period increases in the fourth region we get a new simple structure in the symmetric representation at $\text{Re}(\theta)=(n+\tfrac{1}{2})\tau$ and the fractal structure moves down towards the real $\theta$ line. After the constant solution, the value of $S_\text{sym}$ immediately changes from $-(N-2)/(N+2)$ to $-1$ so that we have the change of colors from light blue to dark pink in figure~\ref{fig_2dcurve}.
\begin{center}
\includegraphics[width=.75\textwidth]{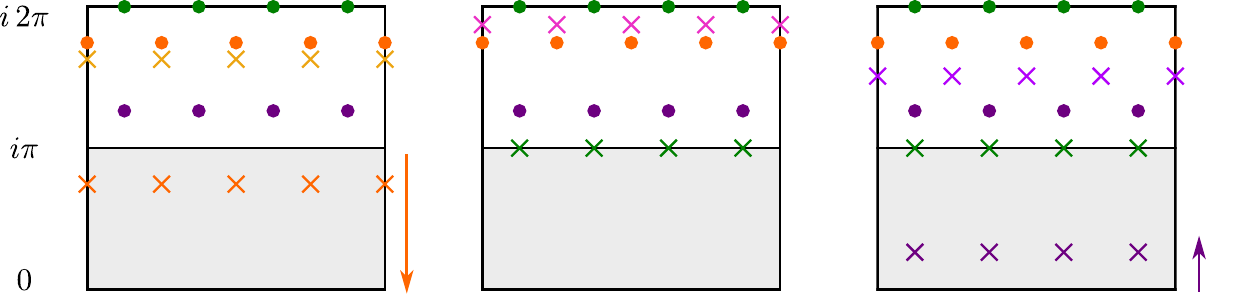}
\end{center}
To reach the final point of (-) free theory, all zeros and poles should disappear and a change of sign in $S_\text{sing}(\theta=0)$ should occur (so that we pass from dark pink to black in the notation of table~\ref{tab:colors}). Most of the structure disappears as the period again diverges. For the change of sign, the fractal structure in the singlet channel reaches $\theta=0$ and collides with its unitarity image pole. Thanks to the fractal structure, similar cancellations occur at $\theta=i\pi n$. Up to an overall minus sign, this leaves us back where we started so by following the same logic we can describe the S-matrices on the lower curve of figure~\ref{fig_2dcurve}.

As we have seen, there are basically three mechanisms for the appearance/disappearance of structures of poles and zeros in the S-matrices. Namely, collision of zeros and poles,  structures moving in the imaginary rapidity direction to higher and higher sheets or moving in the real rapidity direction (e.g. with the period diverging). Although the functions on the 3D monolith are more complicated, the same mechanisms survive.

\subsection{General analytic properties of the Monolith}\la{app_gral3D}
Let us now explore the more general S-matrices on the boundary of the 3D monolith. As one might expect, having a volume with many faces, vertices and edges instead of the $s^*=2m^2$ plane with a single boundary curve significantly adds complexity to the playground. The biggest difference compared to the problem described in the previous section is that the S-matrices on the monolith are not exactly periodic but have a \textit{generalized periodicity}. The explanation for this property --as the periodicity in the 2D plane-- remains an unsolved mystery.

What we mean by the term \textit{generalized periodicity} is that the S-matrices are composed of a central structure with purely imaginary zeros and poles 
and other structures with equally spaced zeros and poles that appear after some offset in real rapidity. In an equation, the S-matrices have the following form:
\beq\la{S_gral3D}
S(\theta)=\mathcal C(\theta) \mathcal G(\theta+\zeta) \mathcal G(-\theta+\zeta)\,,
\eeq
where $\mathcal C(\theta)$ is the central structure, $\zeta$ is the offset and the product $\mathcal G(\theta) \mathcal G(-\theta)$ is periodic in the real rapidity direction.
The first example of such functions was first encountered in \cite{Cordova:2018uop} when studying the S-matrices maximizing the coupling to a single bound state in the singlet channel. Remarkably, a simple modification of this solution describes a line on the boundary of the monolith as described in the next section. For a graphical representation of the type of structure \eqref{S_gral3D}, see figure~\ref{sigma2line}(a).

As far as we can tell numerically, the S-matrices on the boundary of the monolith saturate unitarity except at the constant solution \eqref{constant_soln}. There are only six points where the Yang-Baxter equations are satisfied, corresponding to $\pm$(Free, NLSM, pYB) also present in the 2D plane. 

At a generic point on the boundary, the fractal structures described in the previous section are still present, but we gain many new parameters from the offset in the real rapidity and the ``independent''\footnote{Of course, this is not strictly true since all parameters are related by crossing.} central structure. We have looked at representative points of some of the faces and edges of the monolith so that we have a rough idea of how the interpolation between different faces takes place. Since we do not have yet the complete picture let us for now restrict to one line on the boundary which we know analytically and where the interpolation between two integrable points is precise.


\subsection{The $\sigma_2=0$ line}\la{app_sigma2line}
There is a special line on the boundary of the monolith identified by $\sigma_2(s^*)=0$. For the two-dimensional slate, this condition selects the periodic YB solution where the S-matrices obey $\sigma_2(s)=0$ (i.e. for any value of $s$) implying $S_\text{anti}(s)=-S_\text{sym}(s)$. In the 3D monolith, we have the same situation which greatly simplifies the task of finding an analytic solution.  A very similar problem was introduced in \cite{Cordova:2018uop} when studying the space of S-matrices maximizing the coupling to a single bound state in the singlet representation giving rise to a solution with the generalized periodicity described above. It turns out that the S-matrix of \cite{Cordova:2018uop} times a simple CDD factor which cancels the unwanted poles in the physical strip perfectly describes the $\sigma_2(s^*)=0$ on the monolith. The final expression is\footnote{When comparing equations (37-38) in \cite{Cordova:2018uop} to \eqref{sigma2line}, it is useful to note that 
\beq\nn
F(\lambda ,-\theta ) F(2 \pi -\lambda ,-\theta )=\frac{\sinh (\theta )-i \sin (\lambda )}{\sinh (\theta )+i \sin (\lambda )}\,\frac{+i \theta -\lambda +\pi }{-i \theta -\lambda +\pi }\,F(\pi -\lambda ,\theta )^2\,.
\eeq
}
\beq\la{S_mus}
{\bf{S}}(\theta)=\pm
\begin{pmatrix}
G(\theta)\\
1 \\
-1
\end{pmatrix}
F_{\lambda}(-\theta)F_{2\pi-\lambda}(-\theta)
\[ \prod\limits_{i=1}^{\infty}\frac{\mu_i +i \theta}{\mu_i  -i \theta}\,F^2_{\mu_i }(\theta) \] 
\[ \prod\limits_{n=0}^{\infty}F_{-i\zeta-\frac{in\pi^2}{\nu}}(\theta) F_{i\zeta+\frac{in\pi^2}{\nu}}(\theta) \]\,,
\eeq
where we have defined
\small
\beq\la{G_mus}
G(\theta)\equiv\frac{i\theta-\lambda}{i\theta+\lambda}\,\frac{i\theta+\lambda-2\pi}{i\theta-\lambda+2\pi}\(\prod\limits_{i=1}^\infty\frac{i\theta+\mu_i-\pi}{i\theta-\mu_i+\pi}\,\frac{i\theta-\mu_i-\pi}{i\theta+\mu_i+\pi}\)\frac{\Gamma\[\tfrac{\nu}{\pi^2}(\theta+\zeta-i\pi)\]\Gamma\[\tfrac{\nu}{\pi^2}(-\theta+\zeta+i\pi)\]}{\Gamma\[\tfrac{\nu}{\pi^2}(\theta+\zeta+i\pi)\]\Gamma\[\tfrac{\nu}{\pi^2}(-\theta+\zeta-i\pi)\]}\,.
\eeq
The infinite set of parameters $\mu_i$ can be consistently truncated and determined (along with the offset $\zeta$) using the crossing equations as explained in \cite{Cordova:2018uop} (see appendix A). The factors containing $\lambda$ and $\mu_i$ are part of the central structure $\mathcal C(\theta)$, whereas the product of gamma functions has precisely the form $\mathcal G(\theta+\zeta)\mathcal G(-\theta+\zeta)$ of \eqref{S_gral3D}\footnote{Recall that the function $F_a(\theta)$ puts poles and zeros in the vertical (imaginary $\theta$) direction according to unitarity and crossing. It is a simple exercise to rewrite \eqref{S_mus} in a real periodicity friendly notation as an infinite product of gamma functions akin to the ones in \eqref{G_mus}. }. The analytic structure is depicted in figure~\ref{sigma2line}~({\bf a}). 

\begin{figure}[t!]
\centering
\includegraphics[width=\textwidth]{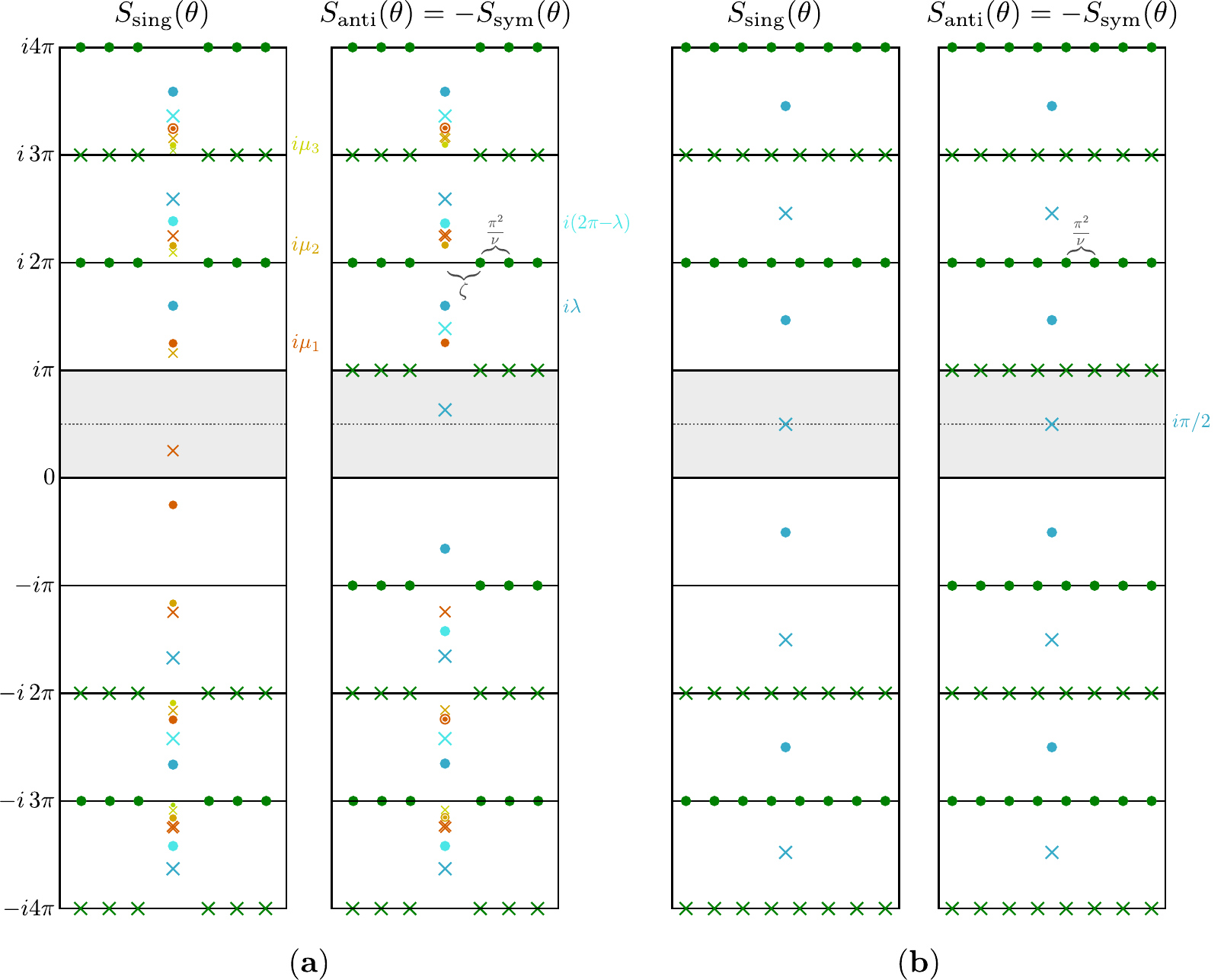}
\caption{({\bf a}) Analytic structure of solution along the $\sigma_2=0$ line given in \eqref{sigma2line}. ({\bf b}) The simple analytic structure remaining from  \eqref{sigma2line} when $\lambda=3\pi/2$.}
\la{sigma2line}
\end{figure}

This solution nicely interpolates between the $\pm$ periodic YB solutions, the two signs referring to the two different lines connecting the integrable solutions. The interpolation takes place as follows. The parameter $\lambda$ --which in \cite{Cordova:2018uop} was related to the mass of the bound state-- takes values $\lambda\in\[\pi,2\pi\]$ so that the first zero in the antisymmetric and symmetric representations remains inside the physical strip (blue cross in figure~\ref{sigma2line}~({\bf a})). It can also be used as a parameter for the position along the two lines. 

As $\boldsymbol{\lambda\rightarrow2\pi}$ three things happen: first, the anti/sym zero in blue reaches the upper boundary of the physical strip; meanwhile, the orange tower of poles and zeros moves down until the zero in the physical strip of the singlet channel arrives to $\theta=0$, producing an infinite cancellation of poles and zeros at $S_a(i\pi n)$; finally the offset reaches the value $\zeta=\pi^2/\nu$ so that we have exactly the analytic structure of the periodic YB solution. 

When $\boldsymbol{\lambda\rightarrow3/2\pi}$ something curious happens: the first anti/sym zero (in blue) moves down to the middle of the physical strip and at the same time the first zero in singlet (orange) moves up also to the middle of the physical strip. Again, infinite cancellations occur, leaving behind a single tower of poles and zeros in the imaginary axis and as $\zeta\rightarrow\pi^2/(2\nu)$ we have the very symmetric solution:
\beqa
{\bf S}(\theta)&=&\Big(N\,\sigma(i\pi-\theta)+\sigma(\theta),\,-\sigma(\theta),\,\sigma(\theta)\Big)\,,\\
\text{with   }\;\;\sigma(\theta)&=&\tan \left(\frac{i \theta }{2}+\frac{\pi }{4}\right)\,\[ \prod\limits_{n=0}^{\infty}F_{-\frac{i\pi^2}{\nu}\(n+\tfrac{1}{2}\)}(\theta) \,F_{\frac{i\pi^2}{\nu}\(n+\tfrac{1}{2}\)}(\theta) \]\,,\nn
\eeqa
whose analytic structure is depicted in figure~\ref{sigma2line}~({\bf b}). On the monolith, this point corresponds to the middle of the green faces in figure~\ref{fig_morph}.

Finally, we have the limit $\boldsymbol{\lambda\rightarrow\pi}$ which leads us to the other periodic YB solution. Here, we have the blue structure going towards $\theta=0$ while the orange one keeps moving up until it reaches the upper boundary of the physical strip. In this case, the offset vanishes $\zeta=0$. Again, the fractal structure of the $\mu$ tower permits the perfect cancellation of poles and zeros so that only the periodic resonances of pYB remain.

As a last remark for this section, let us point out that the fact that we have $S_\text{anti}(s)=-S_\text{sym}(s)$ for any $s$ on this line clarifies the double change of sign in $S_a(\theta=0)$ resulting in the coloring shown in figure~\ref{fig_morph} (from dark (light) green to dark (light)  blue edge where pYB lives). In more generic situations, we expect contiguous colors on the monolith to correspond to a change of sign in a single representation.

\section{Two Mathematica Codes for the Slate}\la{app_periodiccode}

Here we illustrate how to find a very good approximation to the slate in a few seconds. We will work with $O(7)$ symmetry, periodic functions with a small frequency (i.e. large period) $w=1/3$ with $10$ Fourier modes of each sign, $20$ grid points where we impose unitarity inside the fundamental period (in the primal problem) and $100$ grid points used to evaluate the integrals by Chebychev method (in the dual problem) with some high precision. Finally, we solve the dual and original problems at $100$ different points to generate some nice plots. All this translates into the initialization code \\ \\  
\verb"n=7; Nmax=10; gridPoints=20; integralPoints=20; precision=100; plotPts=100; w=1/3;" \\ \\
The crossing matrix is also used in both the dual and the original problem. It is after all where the specific $N$ in $O(N)$ is input. It reads \\ \\
\verb"c={{1/n,1/2-n/2,-1/n+(1+n)/2},{-1/n,1/2,1/2+1/n},{1/n,1/2,1/2-1/n}};" \\ \\ 
We can now set up the primal and dual problems.
\subsection{Primal Problem (Normals)}  \la{OriginalD}
In the primal problem we parametrize crossing symmetric S-matrices. We can use dispersion relations as in \cite{He:2018uxa} and \cite{Cordova:2018uop} or complex plane foliations as in \cite{Paulos:2017fhb} and \cite{EliasMiro:2019kyf}. Here we use a Fourier decomposition and focus on functions with a fixed period. The larger is the period we use the better we approximate a generic function. With the small frequency we chose above we already get a very good approximation to the optimal solution as we will see below. Under crossing positive and negative frequency modes get interchanged so that it is straightforward to write down a crossing symmetric ansatz as \\ \\ 
\verb"S[t_]={sing[0],anti[0],(n*anti[0]+2*sing[0])/(n+2)}+Sum[{sing[n],anti[n],sym[n]}*"\\
\verb"Exp[I*n*t*w]+c.{sing[n],anti[n],sym[n]}*Exp[I*n*(I*Pi-t)*w], {n,1,Nmax}];" \\ \\ 
The free parameters are the various fourier coefficients which we list as \\ \\ 
\verb"vars=Variables[S[1/2]];"\\\\
Finally we prepare the unitarity constraints \\ \\ 
\verb"unit[t_]=ComplexExpand[Re[#1]]^2 + ComplexExpand[Im[#1]]^2 <= 1 & /@ S[t];"\\
\verb"Unitarity=And@@Flatten[unit /@N[Range[0,2*Pi/w,2*Pi/(w*gridPoints)],500]];" \\  \\
and define the components $\sigma_1$ and $\sigma_2$ since we will be plotting the allowed space in this plane. These components are simple combinations of the S-matrix irreps \\ \\
 \verb"{s1,s2} = ({(#1[[3]] - #1[[2]])/2, (#1[[3]] + #1[[2]])/2} & )[S[I*(Pi/2)]]; " \\ \\
Then we have our main function  \\ \\ 
\verb"fsol[alpha_]:=fsol[alpha]=FindMaximum[{Cos[alpha]s1+Sin[alpha]s2,Unitarity},vars]" \\ \\ 
which looks for the slate boundary point with angle $\alpha$ normal. We can run it for many alphas to generate a beautiful plot,\footnote{The \texttt{Dynamic} function with the \texttt{ProgressIndicator} converts an agonizing wait into a bliss. The code works equally well without it but generates more stress in the user.} \\ \\ 
\verb"Dynamic[ProgressIndicator[a,{0,\[Pi]}]]" \\ 
\verb"ListLinePlot[Join[#,-#]&@Table[{s2,s1}/.fsol[a][[2]],{a,0,Pi,Pi/plotPts}],Mesh->All]" \\ \\ 
The reader who runs this  \verb"Mathematica" code should hopefully have obtained the blue dots in figure \ref{AppendixFigure}. The red dots correspond to the dual solution of the next section. Clearly, even with such small parameters and only with a few seconds wait, we can already get a pretty satisfactory approximation to the optimal bounds both from the primal or dual perspective. What is more, we can also directly compare the S-matrices obtained through the original and primal problems using (\ref{dp10}) and indeed obtain a perfect match as another nice confirmation of a zero duality gap as expected.  

\begin{figure}[t]
\centering
\includegraphics[width=.92\textwidth]{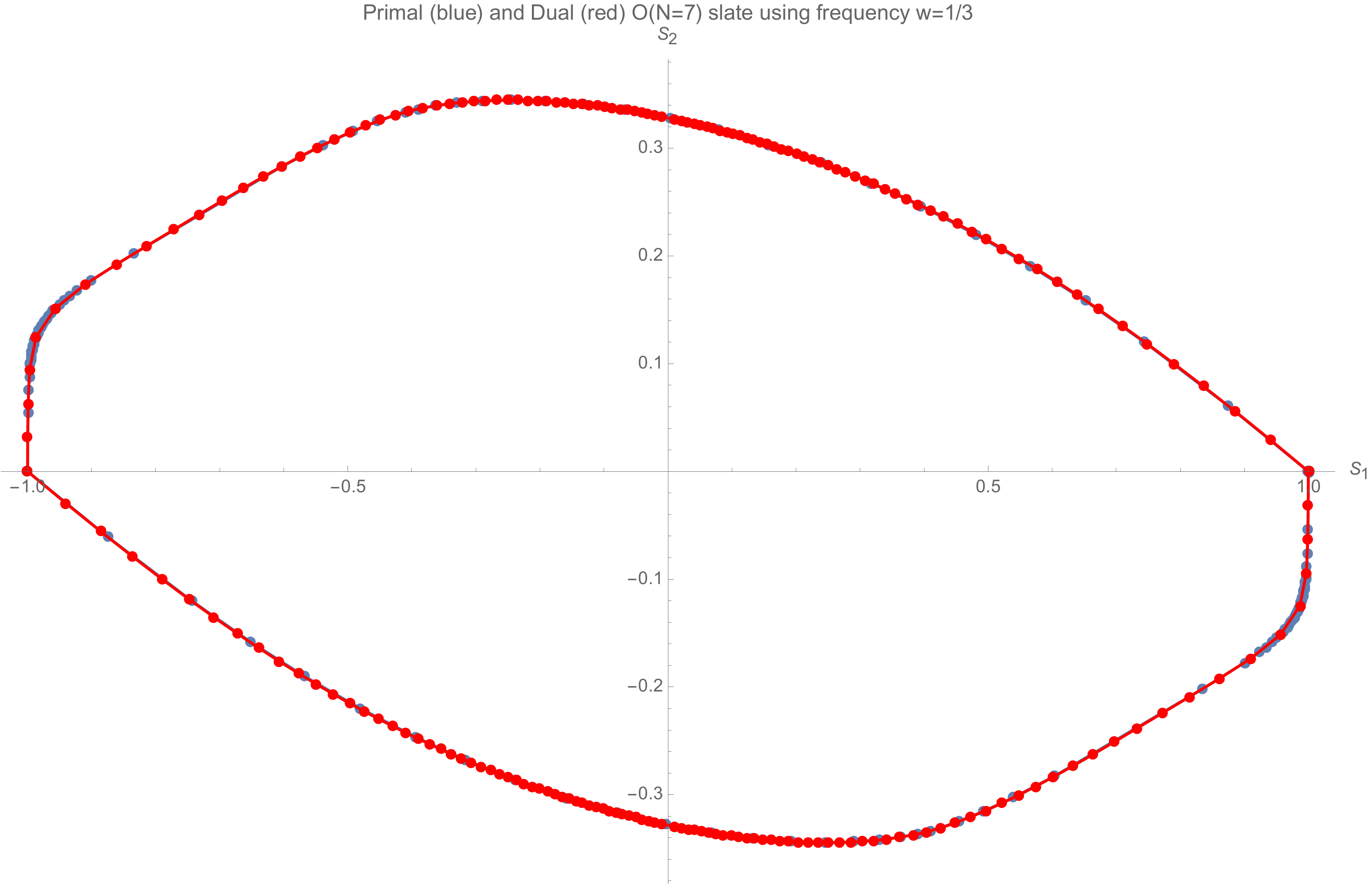}
\caption{Primal (blue) and Dual (red) estimates of the slate boundary with the parameters described in the text. The optimal bound must be somewhere between the two curves; since they are basically on top of each other already, we conclude that the dual and primal problems for this large period are a very good approximation of the optimal bound.
(The blue points are more densely located in larger curvature regions since we used the normal functionals while the red dots are more uniformly distributed since we found them using the radial method.) }
\la{AppendixFigure}
\end{figure}

\subsection{Dual Problem (Radials)} \la{DualD}
In the dual problem we parametrize the kernels $K_a(\theta)$. They have a pole at $i \pi/2$ with residues related to the radial direction (or to the normal) which we want to explore in the slate. It is again straightforward to write down a Fourier ansatz with the right crossing properties:  \\ \\
\verb"v1 = {0, 1/2, 1/2}; v2 = {1/(2*n), -(1/4), (n - 2)/(4*n)};" \\
\verb"K[t_]=(#+c\[Transpose].(#/.t->I*Pi-t)&@Sum[{sing[n],anti[n],sym[n]}*(Exp[I*n*t*w]-Exp[-n*w*Pi/2]),{n,1,Nmax}]+a1*v1+a2*v2) Sech[t];"\\ 
\verb"vars=K[1/2]//Variables" \\ \\
and again we expect the results derived from this ansatz to better approach the optimal slate boundary as we take larger and larger periods. Different $a_1,a_2$ correspond to different directions in the slate; the vectors $v_1$ are the eigenvectors of the transposed crossing matrix with eigenvalue $1$. In the dual problem we don't need to impose any (unitarity) constraints but we do need to compute an integral of the absolute value of the $K_a$ over the real line and then minimize this quantity. For that we write down a very precise evaluation of the integral using Chebychev integrations so that the resulting expression can be minimized using Mathematica's built-in functions. This is achieved through \\ \\ 
\verb"grid=Table[N[Cos[j\[Pi]/(integralPts+1)],precision],{j,1,integralPts}];" \\
\verb"integrals=2*Table[Expand[Times@@(x-Drop[grid, {k}])/Times@@(grid[[k]]-Drop[grid,{k}])]" \\
\verb"/.x^(m_.):>Boole[EvenQ[m]]/(m+1),{k,integralPts}]" \\ 
\verb"f[y_]=(1/2)*Sec[Pi*y/2]^2*Total@Abs@K[Tan[(Pi*y)/2]];" \\
\verb"goal=(f/@grid//ExpandAll//Chop).integrals;"  \\ \\
which produces the integral as \verb"goal" which we simply minimize as\footnote{The \texttt{Quiet} at the end is not very scientific. It quiets \texttt{Mathematica} so we don't see her error message complaints. In this case it is justified since the final results are quite ok and her worries are unjustified. Still, by increasing \texttt{WorkingPrecision} and/or \texttt{PrecisionGoal} one can get rid of such annoyances. The result would be safer but slower so we do not worry about it here.} \\ \\ 
\verb"sol[a_]:=sol[a]=FindMinimum[goal/.First@Solve[Cos[a]*a1+Sin[a]*a2==1],vars]//Quiet" \\ \\
repeating for several radial directions we finally generate a beautiful plot as  \\ \\
\verb"Dynamic[ProgressIndicator[a,{0,\[Pi]}]]" \\
\verb"Table[sol[a][[1]] {Cos[a],Sin[a]},{a,Range[0,\[Pi],\[Pi]/plotPts]}];"\\
\verb"ListLinePlot[Join[%,-%],PlotStyle->Red,Mesh->All]"\\ \\
These are the red dots in figure \ref{AppendixFigure}. Note that we are using the radial constrains (\ref{dp18}) and the relation (\ref{dp20}) to convert the dual problem outcome directly into a statement about the $O(N)$ slate boundary. 



\section{The $O(2)$ Slate}\la{app_N2}

\begin{figure}[t]
\centering
\includegraphics[width=\textwidth]{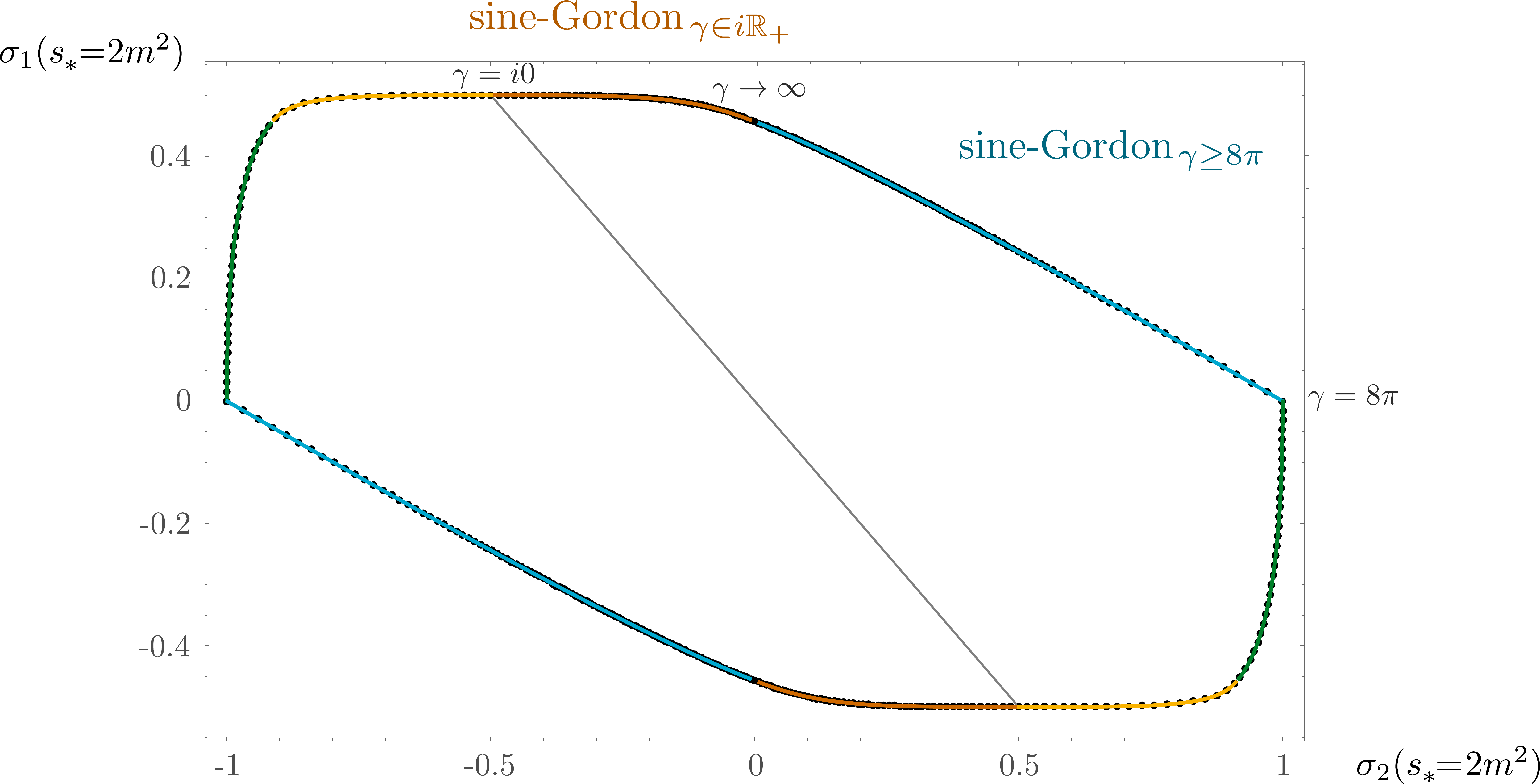}
\caption{The $O(2)$ slate. The black dots are numerical data obtained with the dual minimization explained in appendix~\ref{app_periodiccode} for a small frequency $w=1/8$. The blue/green curves correspond to the analytic solution \eqref{S_sGkinks} with $\gamma\geq8\pi$ whereas the orange/yellow ones are obtained with $\gamma\in i\mathbb R_+$. The blue and green (orange and yellow) sections are related by the map $\sigma_2\rightarrow-\sigma_2-2\sigma_1$, as highlighted by the $\sigma_2=-\sigma_1$ line in grey. }
\la{fig_O2slate}
\end{figure}

The nature of the space of $O(N)$ S-matrices is different for $N>2$ and $N=2$. A simple way to see this is that the integrable solutions for one and other case are completely different. In the former we have the NLSM and periodic Yang-Baxter solutions discussed in the main text, which have no free parameters and therefore stand as isolated points on the boundary of the monolith. In the latter there is an integrable solution with a continuous parameter describing a line on the boundary of the $O(2)$ monolith. In this appendix we focus on the $s_*=2m^2$ slate for $N=2$.

The well known integrable solution for $N=2$ is the sine-Gordon scattering of kinks/antikinks which has a free parameter $\gamma$ related to the coupling in the sine-Gordon Lagrangian. It was first bootstrapped in  \cite{Zamolodchikov:1978xm} and reads
\beq
{\bf S^{sG}_\gamma}(\theta)=-\frac{1}{\pi}\,U(\theta)
\begin{pmatrix}
\sin\(\frac{8\pi i\theta}{\gamma}\)-\sin\(\frac{8\pi^2}{\gamma}\)\\ 
\sin\(\frac{8\pi i\theta}{\gamma}\)+\sin\(\frac{8\pi^2}{\gamma}\)\\ 
-\sin\(\frac{8\pi(\pi+i\theta)}{\gamma}\)\\
\end{pmatrix}\,,\la{S_sGkinks}
\eeq
where again we used the notation $\mathbf S=(S_\text{sing},S_\text{anti},S_\text{sym})^\intercal$ and the prefactor is given by\footnote{
It is sometimes convenient to use the following integral representation for the prefactor:
\beq
U(\theta)=-i \pi\frac{  \sinh (\theta )}{\sin \left(\frac{8 \pi ^2}{\gamma }\right)}\,  
\exp \left\{\frac{1}{2 \pi  i}\int\limits_{-\infty }^{\infty }
 \text{csch} (x-\theta ) \,\log \left[-\frac{2 \sin ^2\left(\frac{8 \pi ^2}{\gamma }\right) \text{csch}^2(x)}{\cos \left(\frac{16 \pi ^2}{\gamma }\right)-\cosh \left(\frac{16 \pi  x}{\gamma }\right)}\right]
  \, dx\right\}\,.
\eeq}
\beqa
U(\theta)&=&\Gamma\(\frac{8\pi}{\gamma}\)\Gamma\(1+i\frac{8\theta}{\gamma}\)\Gamma\(1-\frac{8\pi}{\gamma}-i\frac{8\theta}{\gamma}\)\prod\limits_{n=1}^\infty\frac{R_n(\theta)R_n(i\pi-\theta)}{R_n(0)R_n(i\pi)}\la{U_N2}\,,\\
R_n(\theta)&=&\frac{\Gamma\[2n\frac{8\pi}{\gamma}+i\frac{8\theta}{\gamma}\]\Gamma\[1+2n\frac{8\pi}{\gamma}+i\frac{8\theta}{\gamma}\]}{\Gamma\[(2n+1)\frac{8\pi}{\gamma}+i\frac{8\theta}{\gamma}\]\Gamma\[1+(2n-1)\frac{8\pi}{\gamma}+i\frac{8\theta}{\gamma}\]}\nn\,.
\eeqa

For $\gamma\geq8\pi$ the above S-matrix exhibits no bound states and so our bootstrap problem should make contact with this solution.\footnote{This solution appeared already in the S-matrix bootstrap context \cite{Cordova:2018uop,Paulos:2018fym} in the regime where $\gamma<8\pi$ and there are bound states in the theory.} Amusingly, the whole boundary of the slate can be identified with \eqref{S_sGkinks} and simple modifications of it. 

The results are summarized in figure~\ref{fig_O2slate}. First, we have the blue section which is simply the sine-Gordon S-matrix \eqref{S_sGkinks} with $\gamma\geq8\pi$. The right free theory vertex with $\sigma_2=1$ corresponds to $\gamma=8\pi$ and the point at which $\sigma_2=0$ (which would be the analogue of the periodic YB solution for $N>2$) is reached as $\gamma\rightarrow\infty$. Then we have the orange curve which follows from the same sine-Gordon S-matrix with $\gamma$ purely imaginary $\gamma\in i \mathbb R_+$. Naturally, the $\sigma_2=0$ point connects the two regions at infinity in the $\gamma$ complex plane. These are the two fundamental regions. The rest of the curve can be obtained by the usual reflection $\sigma_i\rightarrow-\sigma_i$ and a map $\sigma_2\rightarrow-\sigma_2-2\sigma_1$ which can be traced back to a simple change of sign in the U(1) basis of the problem.

As a final remark, let us comment that $O(2)$ slate nicely connects to the space of $Z_4$ S-matrices described in \cite{Zamolodchikov:1979ba} and bootstrapped in \cite{susy}. Indeed, by taking two different limits of the integrable elliptic deformation of \cite{Zamolodchikov:1979ba} the two sine-Gordon solutions at the boundary of the $O(2)$ slate ($\gamma\geq8\pi$ and $\gamma\in i \mathbb R_+$) are recovered.\footnote{A special thanks to Alexandre Homrich for dicussions on the relation to the $Z_4$ S-matrix explored in \cite{susy}.}


\bibliographystyle{ieeetr}
\bibliography{ONShape}

\end{document}